\documentclass[aps,prb,amsfonts, amssymb, amsmath, reprint, twocolumn,longbibliography]{revtex4-2}
\usepackage[english]{babel}
\usepackage{graphicx}
\usepackage{amsmath,amssymb,amsthm,mathtools}
\usepackage{physics}
\usepackage{bm}
\usepackage{xcolor}
\usepackage{siunitx}
\usepackage{hyperref}
\hypersetup{
    colorlinks=true,
    citecolor=blue,
    linkcolor=red,
    urlcolor=blue
}

\theoremstyle{plain}
\newtheorem{theorem}{Theorem}[section]
\newtheorem{lemma}[theorem]{Lemma}
\newtheorem{proposition}[theorem]{Proposition}
\newtheorem{corollary}[theorem]{Corollary}
\theoremstyle{definition}

\theoremstyle{remark}

\newcommand*{\hham}{\hat{\mathcal{H}}}
\newcommand*{\Sop}{\hat{\mathbf{S}}}
\newcommand*{\sop}{\hat{\mathbf{s}}}
\newcommand{\ii}{\mathrm{i}}
\newcommand{\ee}{\mathrm{e}}

\newcommand{\Gret}{\mathbf{G}^{R}}
\newcommand{\Gadv}{\mathbf{G}^{A}}
\newcommand{\Gless}{\mathbf{G}^{<}}
\newcommand{\Iodd}{I_{S_z}^{\mathrm{odd}}}
\newcommand{\Ieven}{I_{S_z}^{\mathrm{even}}}

\begin{document}

\title{
Achiral-Chiral Dissociation and Structural Zero of the Current in the Spinterface Model of CISS
}

\author{Seif Alwan}
\affiliation{Materials Theory, Department of Physics and Astronomy, Uppsala
University, Box 516, 75120 Uppsala, Sweden}

\date{\today}

\begin{abstract}
We study the transport sector of the spinterface model of chirality-induced
spin selectivity (CISS) under spin-resolved dissipation, extending the quantum
non-equilibrium Green's function formulation of
Ref.~\cite{Alwan2026QSpinterface} to spin-resolved lead broadenings,
parameterised by a mean level width $\Gamma_0$ and a spin asymmetry
$\Delta\gamma$. Seven results emerge. (i) The spin current decomposes into an
achiral part $\Ieven$, driven by the interfacial moment, and a chiral part
$\Iodd$, driven by the Rashba spin flip. Under $\Delta\gamma$, $\Ieven$ grows
linearly (linear-plus-offset law) while $\Iodd$ is essentially
$\Delta\gamma$-independent, following the geometric-mean suppression.
(ii) The two sectors also dissociate along $\Gamma_0$: the achiral
amplification coefficient, the chiral signal, and the achiral baseline. 
(iii) The coupling optimum $\Gamma_0^*\approx 2.9\,t_{\mathrm{mol}}$ and bias
optimum $V^*=0.90\pm 0.02$~V are operating points of the reference
configuration, not universal scaling laws. 
(iv) A structural zero appears at $\lambda_R\sin(\phi/2)=t_{\mathrm{mol}}$ in
the collinear-lead configuration: $\Iodd$ collapses to machine precision,
universally across chain lengths, frequency resolutions, and twist angles. The
zero originates from the singular $2\times2$ bond matrix, whose null
eigenvector traps an electron in a spin-superposition dark state that does not
hop. 
(v) With non-collinear ferromagnetic leads, the zero is broken in a controlled
way: the residual is generically linear in $P\sin\theta$, arising from a single
coherent spin flip at a boundary, and becomes quadratic at a tuned cancellation
angle where two boundary contributions interfere destructively, with the
exponent jumping sharply from $\beta=1$ to $\beta=2$.
(vi) The chirality-specific signature is a polarization sign reversal at
$\phi\approx 75^\circ$ with amplitude $\approx 6.6\%$. (vii) The global maximum
of $|\Iodd(\Gamma_0,V)|$ 
is at $(1.453~\mathrm{eV}, 0.833~\mathrm{V})$, with
contrast $6\times 10^2$ against the low-bias low-coupling corner. 
These results establish the spinterface model as a transport platform with
a two-axis dissociation structure, non-scaling operating optima, and an exact
structural zero whose non-collinear deformation is a clean
linear-plus-quadratic hierarchy.
\end{abstract}
\maketitle

\section{Introduction}
\label{sec:intro}

The chirality-induced spin selectivity (CISS) effect is the observation that
electrons transmitted through a chiral medium acquire a net spin polarisation
tied to the medium's handedness. It has been reported across a remarkable range
of systems: from helical peptides and DNA~\cite{Naaman2019,Gohler11,Evers22},
to single-molecule junctions~\cite{Safari2024}, chiral
crystals~\cite{Bloom2024ChemRev}, and hybrid organic-inorganic
perovskites~\cite{Dong2025AnnuRev}. Since the pioneering experiments of Ray
\emph{et al.}~\cite{Ray99} and G\"ohler \emph{et al.}~\cite{Gohler11}, the
effect has been harnessed for applications in spin-based memory, chiral
catalysis, and enantioselective separation. Comprehensive reviews now cover the
breadth of molecules and materials that exhibit CISS-based
phenomena~\cite{Evers22,Naaman2019,Bloom2024ChemRev}.

Two broad classes of explanations have been proposed. The first is
\emph{structural}: an effective spin-orbit field that arises from the
helical geometry of the molecule and the resulting chiral electrostatic
potential~\cite{Gutierrez2012,Shitade2020,Michaeli2019}. The second is
\emph{interfacial}: a spin-dependent hybridisation between the molecule
and a magnetic or spin-active electrode, often referred to as the
spinterface mechanism~\cite{Alwan2021,Dubi2022}. These two classes are
not mutually exclusive, and are increasingly recognised as complementary
perspectives on a single underlying interface-mediated mechanism; the
spinterface model studied in this paper shows how they can coexist and
cooperate.

A recent critical perspective by Sarkar \emph{et al.}~\cite{Sarkar2025}
provides a comprehensive overview of the spinterface mechanism, which
posits a feedback interaction between electron motion in chiral molecules
and fluctuating surface magnetic moments. While the spinterface model has
been shown to quantitatively reproduce experimental data across various
systems and conditions~\cite{Sarkar2025}, a fully quantum formulation of
the mechanism has been lacking. The perspective paper explicitly calls
for a rigorous theoretical framework to address open questions, including
the nature of surface magnetism, the role of dissipation, and the
applicability of the mechanism to non-helical or electrode-free systems.
This work directly addresses that call.

The fully quantum non-equilibrium Green's function (NEGF) formulation of
the spinterface model was established in Ref.~\cite{Alwan2026QSpinterface},
where it was shown that the spin current decomposes into an \emph{even}
part under chirality reversal,
\begin{equation}
\Ieven(\phi) = \tfrac12\bigl[I_{S_z}(+\phi) + I_{S_z}(-\phi)\bigr],
\label{eq:Ieven_def}
\end{equation}
and an \emph{odd} part,
\begin{equation}
\Iodd(\phi) = \tfrac12\bigl[I_{S_z}(+\phi) - I_{S_z}(-\phi)\bigr],
\label{eq:Iodd_def}
\end{equation}
with $\Ieven$ reflecting the achiral spin-valve baseline from the
interfacial moment and $\Iodd$ the genuine chiral contribution.
Reference~\cite{Alwan2026QSpinterface} further showed that $\Iodd$
vanishes identically at fixed moment for chains of fewer than four
total sites (one gold, three molecular), establishing a geometric
threshold for the CISS mechanism.

Throughout this paper, "dissipation" refers specifically to
spin-dependent lead broadening, characterized by the gold-level channel
$\Delta\gamma_{\mathrm{gold}}$ of Ref.~\cite{Alwan2026EP}; the
chain-level channel $\Delta\gamma_{\mathrm{chain}}$ is not studied
here, nor is it the electron-phonon dissipation central to Fransson's
mechanism~\cite{Fransson2020PRB,Fransson2023ChiralPhonon}. The
conclusions are therefore scoped to this channel, and we return to the
comparison with the microscopic dissipation mechanism in
Sec.~\ref{sec:dissociation}.

To make the present paper self-contained, we recall the essential
structure of the model. The junction consists of a gold atom (site
$0$) carrying a self-consistent mean-field moment $m_z$ at the
electrode-molecule interface, a chiral molecular chain of $L$ sites
(sites $1,\ldots,L$), and two wide-band non-magnetic leads. The
chirality of the molecule is encoded in a Rashba spin-flip hopping
$\ii\lambda_R\sin(\phi/2)$ per bond, which cannot be gauged away and is
the microscopic origin of chirality-dependent spin filtering. An exact
antiunitary symmetry $H^*(\phi) = H(-\phi)$ of the Hamiltonian implies
a reciprocity relation $|G^R_\phi[i,j]| = |G^R_{-\phi}[j,i]|$ for the
retarded Green's function. In a device that is left-right reciprocal
in the magnetic sense this relation would force $T_\sigma(\phi) =
T_\sigma(-\phi)$, and hence $\Iodd = 0$ for all chains. The
spinterface junction is \emph{not} left-right reciprocal: the magnetic
moment sits on the gold end and not on the molecular end. The odd part
is therefore permitted by symmetry and arises generically for
$L \geq 3$, which is the smallest chain length at which two or more
Rashba-active bonds are present to defeat the reciprocity of the
transmission magnitudes.

In Ref.~\cite{Alwan2026EP}, we analysed the non-Hermitian structure of the
effective Hamiltonian $H_{\mathrm{eff}}=H_{\mathrm{MF}}+\Sigma_L^R+\Sigma_R^R$
and showed that the model admits two algebraically distinct
near-exceptional-point (EP) regimes. A \emph{gold-level} near-EP (Scheme~B)
requires spin-symmetric lead broadening $\Delta\gamma_{\mathrm{gold}}=0$ and an
anti-Hermitian, electrode-mediated spin flip. A \emph{chain-level} near-EP
(Scheme~A) requires spin-dependent chain broadening
$\Delta\gamma_{\mathrm{chain}}\neq 0$ and a Hermitian on-site Rashba spin flip.
Mixing the diagonal cancellation of one scheme with the off-diagonal structure
of the other yields a strictly negative discriminant and admits no EP; this
makes the gold/chain classification exhaustive.

Beyond the structural and interfacial routes discussed above, recent years have
seen substantial progress in several related directions that are directly
relevant to the present work. The CISS effect has been extended from molecular
junctions to chiral crystals, hybrid organic-inorganic perovskites, and
supramolecular assemblies~\cite{Bloom2024ChemRev,Dong2025AnnuRev}. In parallel,
chiral phonons have emerged as a new route to spin polarization, in which
angular momentum carried by chiral lattice vibrations is transferred to
electronic spins~\cite{Fransson2023ChiralPhonon,Juraschek2025ChiralPhonons},
and this mechanism has been shown to enhance ferromagnetism in hybrid
systems~\cite{Fransson2025ChiralPhononsFerro}. Experiments have demonstrated
chirality-induced phonon spin selectivity via elastic spin-orbit
interaction~\cite{Yang2024PNASChiralPhonon}. Non-Hermitian physics has also
entered the CISS literature: the non-Hermitian skin effect has been proposed as
the origin of anomalous magnetoresistance in chiral
molecules~\cite{Zhao2025NHSE}; non-Hermitian exchange has been identified as a
unifying origin of CISS~\cite{Theiler2026SciAdv}; and exceptional points have
been shown to be enantiosensitive in open chiral
systems~\cite{Mayer2026EnantioEP} and have been realized in single-molecule
junctions~\cite{Yang2024SpiroEP}. These developments reinforce the central
premise of the spinterface model: spin selectivity in chiral systems is not
confined to a single microscopic channel but can arise from the interplay of
structural chirality, interfacial magnetism, and angular-momentum transfer. The
present work bridges these threads by providing a fully quantum, NEGF-based
formulation that can incorporate such mechanisms on equal footing and by
introducing a diagnostic for the even and odd decomposition of the spin current
that can distinguish genuine chiral signatures from achiral spin-valve effects.

The present paper addresses a question that is complementary to both
predecessors: \emph{how does the transport sector of the spinterface
model respond to dissipation in general, and to spin-dependent
dissipation in particular?} The question is motivated by three
considerations.

\emph{First}, the critical perspective of Ref.~\cite{Sarkar2025}
identifies the role of dissipation as an open problem. Existing
treatments include dissipation as either an unavoidable broadening or as
a qualitative time-reversal-symmetry-breaking ingredient, but no
systematic study has mapped the response of the CISS signal across the
coherent-to-overdamped crossover.

\emph{Second}, the EP taxonomy of Ref.~\cite{Alwan2026EP} predicts that
spin-dependent lead broadening $\Delta\gamma_{\mathrm{gold}}\neq 0$
explicitly \emph{forbids} the gold-level Scheme~B EP. If this prediction
is correct, spin-dependent dissipation should \emph{not} amplify the
chiral spin current at the gold level; the EP mechanism for chiral
amplification is unavailable in this regime. The transport sector should
therefore show the achiral and chiral parts responding differently.

\emph{Third}, the CISS signal has sharp interior optima in the
$(\Gamma_0,V)$ plane, which are \emph{operating points} of the reference
parameter set rather than universal scaling laws. These optima coexist
with one exact symmetry of the model, a structural zero. The structural
zero and the non-collinear deformation of the leads are the subject of
Secs.~\ref{sec:zero} and~\ref{sec:noncollinear}, and they establish the
spinterface model as a platform for exact symmetry results beyond the
CISS mechanism itself.

The dissipation channel studied in this paper is fundamentally different
from the dissipation mechanism developed by
Fransson~\cite{Fransson2020PRB,Fransson2023ChiralPhonon}. In Fransson's
approach, dissipation arises from electron-phonon coupling: the coupling
of the transported electron to molecular vibrations generates a
phonon-mediated self energy that (i) enhances the effective spin-orbit
interaction and (ii) provides the time-reversal-symmetry breaking
required for a stationary spin polarization~\cite{Das2022}. A recent
paper~\cite{Upadhyay2026} has argued that weak electron-phonon coupling
is insufficient to generate significant CISS. That debate concerns the
phonon channel and does not affect our conclusions about the lead
channel. The two approaches are therefore complementary rather than
competing: Fransson's mechanism addresses the \emph{origin} of CISS
through phonon-mediated processes, while the present work addresses the
\emph{transport consequences} of spin-dependent dissipation for a given
interface moment.

To clarify this further: dissipation is a necessary condition for the
chiral signal, it provides the finite lifetime that makes the interference
term nonzero, but is not a sufficient or amplifying condition. The
driving force is the coherent interference between spin-conserving and
spin-flip amplitudes, enabled by the magnetic interface and the chiral
hopping. Dissipation asymmetry $\Delta\gamma$ drives the achiral sector
and works against the chiral sector.

The paper is organized as follows. Section~\ref{sec:model} reviews the
model and specifies the spin-resolved broadening extension.
Section~\ref{sec:methods} describes the numerical methodology.
Section~\ref{sec:dissociation} presents the achiral-chiral dissociation,
including its microscopic origin (arithmetic vs geometric mean of the
lead couplings) and its two-axis structure. Section~\ref{sec:coupling}
presents the coupling-scale optimum and its limits.
Section~\ref{sec:bias} presents the bias optimum and its independence
from $t_{\mathrm{mol}}$. Section~\ref{sec:zero} presents the structural
zero at $k=t$, its universality in $L$, $N_\omega$ and $\phi$, and the
full Green's function proof of its existence.
Section~\ref{sec:noncollinear} presents the non-collinear extension: the
linear and quadratic scaling hierarchy of the structural zero under
rotated ferromagnetic leads. Section~\ref{sec:sign} presents the
chirality sign reversal. Section~\ref{sec:conclusion} concludes. Four
appendices cover numerical convergence, self-consistency conventions, the
choice of the broadening floor, and alternative choices of lead
asymmetry.

\section{Model and Methods}
\label{sec:model}

\subsection{Physical picture and the full Hamiltonian}

The spinterface model attributes CISS to the cooperative action of a
magnetic moment localised at the electrode-molecule interface and the
chiral Rashba hopping along the molecular backbone. In the quantum
formulation of Ref.~\cite{Alwan2026QSpinterface}, the interface is
represented by a gold atom carrying a self-consistent mean-field moment
$m_z$, the molecule by a nearest-neighbour chain with a Rashba
spin-flip amplitude per bond, and the electrodes by two wide-band
leads. The present paper uses this Hamiltonian unchanged; the only
extension is that the lead broadenings are allowed to depend on spin.
We write the full Hamiltonian explicitly so that the paper is
self-contained.

The total Hamiltonian is
\begin{equation}
\hham = \hham_{\mathrm{mol}} + \hham_{\mathrm{Au}} + \hham_{\mathrm{hop}}
+ \hham_{\mathrm{leads}} + \hham_{\mathrm{tun}} + \hham_{\mathrm{exch}},
\label{eq:Htot}
\end{equation}
comprising a chiral molecular chain of $L$ sites
(sites $1,\ldots,L$), a gold atom (site~$A\equiv 0$), a direct
gold-molecule hopping, two non-magnetic leads, and an exchange
coupling that generates the self-consistent moment. The site-spin
basis is
$\{|A\uparrow\rangle, |A\downarrow\rangle, |M_1\uparrow\rangle, \ldots,
|M_L\downarrow\rangle\}$, of dimension $2(L+1)$. The individual terms
are as follows.

\emph{Molecular chain.}
\begin{equation}
\hham_{\mathrm{mol}}
=
\sum_{n,\sigma} \varepsilon_{n\sigma} d_{n\sigma}^\dagger d_{n\sigma}
-
\sum_{\langle n,n'\rangle,\sigma} t_{\mathrm{mol}}\,
d_{n\sigma}^\dagger d_{n'\sigma}
+ \mathrm{h.c.},
\label{eq:Hmol}
\end{equation}
with nearest-neighbour hopping $t_{\mathrm{mol}}$ and site energies
$\varepsilon_{n\sigma}$ given below in Eq.~\eqref{eq:eps_M_sigma}.

\emph{Gold atom.}
\begin{equation}
\hham_{\mathrm{Au}} = \sum_{\sigma} \varepsilon_{A\sigma}\,
d_{A\sigma}^\dagger d_{A\sigma},
\label{eq:HAu}
\end{equation}
with the spin-split on-site energies
$\varepsilon_{A\sigma}$ given below in Eq.~\eqref{eq:eps_A_sigma}.

\emph{Gold-molecule hopping.}
\begin{equation}
\hham_{\mathrm{hop}}
=
-t_{AB}\sum_\sigma\left(d_{A\sigma}^\dagger d_{1\sigma}
+ d_{1\sigma}^\dagger d_{A\sigma}\right).
\label{eq:Hhop}
\end{equation}
The gold-molecule bond is spin-diagonal: only the $L-1$
\emph{molecular}-\emph{molecular} bonds carry a Rashba spin-flip
term.

\emph{Leads.}
\begin{equation}
\hham_{\mathrm{leads}}
=
\sum_{\alpha\in\{L,R\}}\sum_{k,\sigma}
\varepsilon_{\alpha k\sigma}\, c_{\alpha k\sigma}^\dagger c_{\alpha k\sigma},
\label{eq:Hleads}
\end{equation}
with $\varepsilon_{\alpha k\uparrow}=\varepsilon_{\alpha k\downarrow}$
(non-magnetic leads).

\emph{Tunnelling.}
\begin{equation}
\hham_{\mathrm{tun}}
=
\sum_{k,\sigma}\left(V_L\, c_{Lk\sigma}^\dagger d_{A\sigma}
+ \mathrm{h.c.}\right)
+
\sum_{k,\sigma}\left(V_R\, c_{Rk\sigma}^\dagger d_{M_L\sigma}
+ \mathrm{h.c.}\right),
\label{eq:Htun}
\end{equation}
with momentum-independent $V_L$, $V_R$ in the wide-band limit used
below.

\emph{Exchange / spin-orbit coupling.}
\begin{equation}
\begin{split}
\hham_{\mathrm{exch}}
&=
J_{el}\,\Sop_A\cdot\sop_A
+
J'_p\,\hat S_A^z(\hat n_{A\uparrow} - \hat n_{A\downarrow})\\
&\quad+
K\,\hat S_A^z\sum_\sigma\left(d_{A\sigma}^\dagger d_{B\sigma}
+ \mathrm{h.c.}\right),
\label{eq:Hexch}
\end{split}
\end{equation}
where $\sop_A = \tfrac12\sum_{\sigma\sigma'}d_{A\sigma}^\dagger
\bm\sigma_{\sigma\sigma'} d_{A\sigma'}$ is the electronic spin density
at the gold site and $B\equiv M_L$ is the last molecular site. The
three terms represent the direct exchange, the spin-dependent charge
redistribution, and the hybridisation contribution respectively.
The hybridisation term $K$ couples the gold moment to the
gold-molecule charge transfer; it is a mean-field device for
generating an effective field and is not intended to be a physical
long-range interaction.

\emph{On-site energies.}
The mean-field on-site energies are
\begin{equation}
\varepsilon_{A\sigma} = \varepsilon_A
+ \sigma\left(\tfrac{J_{\mathrm{el}}}{2} + J'_p\right)m_z,
\label{eq:eps_A_sigma}
\end{equation}
and
\begin{equation}
\varepsilon_{M\sigma} = \varepsilon_{\mathrm{mol}}
+ \sigma(J'_p + \alpha_{AB})m_z,
\label{eq:eps_M_sigma}
\end{equation}
where $\alpha_{AB}$ is a phenomenological spin-splitting coefficient
for the molecular sites (distinct from the feedback coefficients
$\alpha_A$, $\alpha_1$, $\alpha_K$ defined below, which act only at
the gold site). After the mean-field decoupling
$\Sop_A\to m_z\,\hat z$, the gold block of the Hamiltonian becomes
\begin{equation}
\begin{split}
\hham_{\mathrm{Au}}^{\mathrm{MF}}
&=
\sum_{\sigma\sigma'}
d_{A\sigma}^\dagger
\left[\varepsilon_A\delta_{\sigma\sigma'}
+
\tfrac{J_{el}}{2}(\hat z\cdot\bm\sigma)_{\sigma\sigma'}
+
J'_p m_z\,\sigma^z_{\sigma\sigma'}\right] d_{A\sigma'}\\
&\quad+
K m_z\sum_\sigma\left(d_{A\sigma}^\dagger d_{B\sigma}
+ \mathrm{h.c.}\right),
\label{eq:HAuMF}
\end{split}
\end{equation}
which reproduces Eqs.~\eqref{eq:eps_A_sigma}-\eqref{eq:eps_M_sigma}
upon collecting the diagonal terms. The single-particle mean-field
Hamiltonian $\mathbf H_{\mathrm{MF}}$ appearing in the Green's
function below is the matrix representation of
$\hham_{\mathrm{mol}}+\hham_{\mathrm{Au}}^{\mathrm{MF}}+\hham_{\mathrm{hop}}$
in the site-spin basis.

\subsection{The chiral bond matrix and its singular structure}
\label{sec:bond_matrix}

The chiral nature of the molecule is encoded entirely in the hopping
between nearest-neighbour molecular sites. In the spin block
$\{|M_n\uparrow\rangle, |M_n\downarrow\rangle\}$ it is the
$2\times 2$ matrix
\begin{equation}
  \begin{split}
    \mathbf h &=
\begin{pmatrix}
-t_{\mathrm{mol}}\,\ee^{-\ii\phi/2}
& \phantom{-}\ii\,\lambda_R\sin(\phi/2) \\
-\ii\,\lambda_R\sin(\phi/2)
& -t_{\mathrm{mol}}\,\ee^{+\ii\phi/2}
\end{pmatrix}\\
&=
\begin{pmatrix}
-t\,\ee^{-\ii\phi/2} & \phantom{-}\ii k \\
-\ii k & -t\,\ee^{+\ii\phi/2}
\end{pmatrix},
\label{eq:bond_matrix}
  \end{split}
\end{equation}
where $t\equiv t_{\mathrm{mol}}$ and we introduce the shorthand
\begin{equation}
\begin{split}
k \equiv \lambda_R\sin(\phi/2).
\label{eq:k_def}
\end{split}
\end{equation}
The matrix $\mathbf h$ is Hermitian for real $\phi$ and $k$. Its
characteristic polynomial is
\begin{equation}
\det(\mathbf h - \lambda\mathbf 1)
= \lambda^2 + 2t\cos(\phi/2)\,\lambda + t^2 - k^2,
\label{eq:bond_charpoly}
\end{equation}
so the eigenvalues are
\begin{equation}
\lambda_\pm = -t\cos(\phi/2) \pm \sqrt{k^2 - t^2\sin^2(\phi/2)},
\label{eq:bond_eigenvalues}
\end{equation}
and the determinant is
\begin{equation}
\det\mathbf h = t^2 - k^2.
\label{eq:bond_det}
\end{equation}

\emph{Two distinct singular limits.} The spectrum has two distinguished
parameter values.
At $k = t\sin(\phi/2)$ the two eigenvalues coincide,
$\lambda_+ = \lambda_- = -t\cos(\phi/2)$: since $\mathbf h$ is
Hermitian, this is an ordinary level crossing, not an exceptional
point, and it has no special consequence for transport.
At $k = \pm t$ the determinant vanishes while one eigenvalue remains
finite: $\lambda_+ = 0$, $\lambda_- = -2t\cos(\phi/2)$. The matrix
has \emph{rank 1}, and it admits the exact outer-product
factorization
\begin{equation}
\mathbf h = |u\rangle\langle v|,
\qquad
|u\rangle = \begin{pmatrix}1\\ \ii\,\ee^{+\ii\phi/2}\end{pmatrix},
\quad
\langle v| = \bigl(-t\,\ee^{-\ii\phi/2},\;\ii t\bigr).
\label{eq:rank1}
\end{equation}
The right null vector is
$|w\rangle = (1,\, -\ii\,\ee^{-\ii\phi/2})^T$, and the left null
vector is $\langle\tilde w| = (-\ii\,\ee^{+\ii\phi/2},\, 1)$. The
non-zero (bright) eigenvalue is
$\lambda_{\mathrm{bright}} = \langle v|u\rangle = -2t\cos(\phi/2)$.

\emph{Physical interpretation.} The rank-1 structure at
$k=\pm t_{\mathrm{mol}}$ splits the spin space into a \emph{dark channel}
$|w\rangle$ (kernel of $\mathbf h$) and a \emph{bright channel} $|u\rangle$
(image of $\mathbf h$). An electron in the dark spin superposition cannot hop
across the bond; the bright state hops with amplitude
$\lambda_{\mathrm{bright}}$, which is $\phi$-even. This bond-local rank
deficiency is the algebraic origin of the exact zero of the odd spin current at
$\lambda_R\sin(\phi/2)=t_{\mathrm{mol}}$ that is the subject of
Sec.~\ref{sec:zero}. The rank-1 structure is a \emph{Hermitian} rank
deficiency, categorically distinct from the exceptional points of
Ref.~\cite{Alwan2026EP}, which require non-Hermitian eigenvector coalescence of
the full $N\times N$ effective Hamiltonian.

\emph{Gauge freedom.} The spin-diagonal phases $e^{\pm\ii\phi/2}$
appearing in the diagonal of $\mathbf h$ are pure gauge: they can be
removed by the site-dependent spin rotation
$d_{n\sigma}\to e^{\ii\sigma n\phi/2} d_{n\sigma}$, and they have no
effect on any transport observable. The Rashba amplitude
$\ii\lambda_R\sin(\phi/2)$ is the only gauge-invariant off-diagonal
element and is therefore the microscopic origin of chirality-dependent
spin filtering.

\subsection{Mean-field decoupling and self-consistency}
\label{sec:selfcons}

The magnetic moment at the gold site,
$\mathbf m = \langle\Sop_A\rangle$, is determined self-consistently
from the Brillouin condition
\begin{equation}
m_z = \tfrac12\tanh\!\left(\frac{\beta B_{\mathrm{eff}}}{2}\right),
\label{eq:selfcons_bisection}
\end{equation}
with the effective field
\begin{equation}
B_{\mathrm{eff}} = \alpha_A\langle s_A^z\rangle
+ \alpha_1\Delta s + \alpha_K\langle\hat{\mathcal N}\rangle
+ b_{\mathrm{static}},
\label{eq:Beff}
\end{equation}
where $\Delta s = \langle n_{A\uparrow}\rangle - \langle n_{A\downarrow}\rangle$
is the spin imbalance at the gold site and
$\hat{\mathcal N} = \sum_\sigma(d_{A\sigma}^\dagger d_{B\sigma}
+ \mathrm{h.c.})$ is the hybridisation operator. We work in the
mean-field approximation with $m_z$ the only non-zero component; the
transverse components $m_x$, $m_y$ are set to zero because the
effective field is assumed to point along $\hat z$.

\emph{The three feedback channels.}
(i)~$\alpha_A\langle s_A^z\rangle$ is the direct exchange coupling
between the local moment and the conduction electrons at the gold
site; in a Stoner-like picture it is the exchange energy of a
magnetic impurity in a metal.
(ii)~$\alpha_1\Delta s$ is a charge-redistribution term: when the
gold site accumulates more up-spin than down-spin electrons, the
difference acts back on the effective field via the charging energy.
(iii)~$\alpha_K\langle\hat{\mathcal N}\rangle$ is the hybridisation
contribution: the electrons that hop between the gold site and the
molecule carry spin information, and their contribution to
$B_{\mathrm{eff}}$ depends on the spin character of the hybridised
orbitals. This term is the one that couples the molecular chirality
to the gold moment (through the spin-dependent transmission of the
hybridised states) and is therefore the most sensitive to $\phi$.

\emph{Static field.} The last term $b_{\mathrm{static}}$ is a small
symmetry-breaking field along $\hat z$ (of order $5~\mathrm{meV}$ in
our numerical work) that stabilises the moment against thermal and
quantum fluctuations. Without it, the self-consistent solution at
$T=300~\mathrm{K}$ would be trivially zero by symmetry. The field is
chosen so that the resulting moment
($m_z\approx 0.4$-$0.5$ in the reference configuration) is
determined predominantly by the electronic feedback rather than by
the field itself: for $b_{\mathrm{static}}$ alone,
$m_z = \tfrac12\tanh(b_{\mathrm{static}}/2k_BT)\approx 0.048$ at
$300~\mathrm{K}$, an order of magnitude smaller than the observed
value.

\emph{Electronic averages.}
The expectation values appearing in $B_{\mathrm{eff}}$ are computed
from the lesser Green's function,
\begin{align}
\langle n_{A\sigma}\rangle
&= -\ii\int\frac{d\omega}{2\pi}\,G^<_{A\sigma,A\sigma}(\omega),
\label{eq:nA_def}\\
\langle s_A^z\rangle
&= \tfrac12\bigl(\langle n_{A\uparrow}\rangle - \langle n_{A\downarrow}\rangle\bigr),
\label{eq:sAz_def}\\
\langle\hat{\mathcal N}\rangle
&= -\ii\sum_\sigma\int\frac{d\omega}{2\pi}
\bigl[G^<_{B\sigma,A\sigma}(\omega) + G^<_{A\sigma,B\sigma}(\omega)\bigr],
\label{eq:hyb_def}
\end{align}
with $\Gless = \Gret\Sigma^<\Gadv$ [Eq.~\eqref{eq:Keldysh} below].

\emph{Positive feedback and damping.}
When $\alpha_A$ and $\alpha_1$ are positive, the loop has a
positive-feedback character: an increase in $\langle s_A^z\rangle$
increases $B_{\mathrm{eff}}$, which increases $m_z$, which in turn
increases the spin splitting and hence $\langle s_A^z\rangle$. The
loop is stabilised by the saturation of the Brillouin function at
$m_z\to 1/2$. We use a damped fixed-point iteration,
$m_z^{(k+1)} = \eta f(B_{\mathrm{eff}}^{(k)}) + (1-\eta) m_z^{(k)}$,
with damping $\eta=0.9$ in the room-temperature regime. The
convergence criterion is
$|m_z^{(k+1)} - m_z^{(k)}| < 10^{-7}$ on the undamped step, and
typical convergence requires 20-200 iterations depending on the
bias and the strength of the feedback.

\subsection{Spin-resolved leads}
\label{sec:spin_leads}

The essential extension in this paper is that the lead-molecule
couplings are spin-dependent. In the wide-band limit the retarded
self-energies are
\begin{equation}
\Sigma_\alpha^R = -\frac{\ii}{2}\Gamma_{\alpha}\mathbf P_\alpha,
\label{eq:sigma_general}
\end{equation}
with $\Gamma_\alpha$ spin-resolved:
\begin{equation}
\Gamma_{\alpha,\uparrow} = \Gamma_0 + \frac{\Delta\gamma}{2},
\qquad
\Gamma_{\alpha,\downarrow} = \Gamma_0 - \frac{\Delta\gamma}{2},
\label{eq:spin_resolved_Gamma}
\end{equation}
for $\alpha=L,R$, with $\Gamma_0 > 0$ the mean level width and $\Delta\gamma$
the spin asymmetry. We impose a lower bound
$\Gamma_{\alpha,\sigma}\geq\Gamma_{\mathrm{floor}}=0.02~\mathrm{eV}$ so that
neither spin channel closes completely; the value is chosen to preserve the
physics of an open channel while avoiding the degenerate regime discussed in
Appendix~\ref{app:floor}. Throughout this paper,  we apply the spin asymmetry
symmetrically to both leads.
This is one
of several physically motivated choices; alternatives are discussed in
Appendix~\ref{app:asymmetry_alternatives}.

\emph{Physical origin.} Spin-dependent lead broadenings arise when
the electrode material possesses an intrinsic spin polarisation
(e.g., a magnetic or half-metallic contact) or when the molecule is
coupled to the lead through a spin-selective buffer. In the language
of Ref.~\cite{Alwan2026EP}, this is precisely the gold-level
dissipation channel $\Delta\gamma_{\mathrm{gold}}$, not the
chain-level $\Delta\gamma_{\mathrm{chain}}$. The chain-level spin
asymmetry would require spin-dependent broadening of the chain sites
(not the leads), which would be the physical signature of a
spin-polarised substrate; that regime is not studied here.
Throughout this paper $\Delta\gamma$ denotes
$\Delta\gamma_{\mathrm{gold}}$; the chain-level channel
$\Delta\gamma_{\mathrm{chain}}$ of Ref.~\cite{Alwan2026EP} is not
addressed.

\emph{Wide-band justification.} The wide-band limit is justified when the leads
have a smooth density of states over the energy window $[-2,2]~\mathrm{eV}$
around the Fermi level, which is the case for metallic gold electrodes. 
We have checked that the results reported below are insensitive to a finite
bandwidth of $\pm 5~\mathrm{eV}$. The limit plays a further role in this paper:
the exact structural zero of Sec.~\ref{sec:zero} requires $|G^R|=|G^A|$ at the
boundary, which holds in the wide-band limit and becomes approximate with
corrections of $O((\mathrm{Re}\Sigma^R/\Gamma)^2)$ beyond it.

\subsection{Transport, the even-odd decomposition, and reciprocity}
\label{sec:transport}

The retarded Green's function of the central region is
\begin{equation}
\Gret(\omega) = \left[\omega\mathbf I - \mathbf H_{\mathrm{MF}}
- \Sigma_L^R - \Sigma_R^R\right]^{-1},
\label{eq:Gret}
\end{equation}
the lesser Green's function follows from the Keldysh equation
\begin{equation}
\Gless(\omega) = \Gret(\omega)\,\Sigma^<(\omega)\,\Gadv(\omega),
\qquad
\Sigma^< = \Sigma_L^< + \Sigma_R^<,
\label{eq:Keldysh}
\end{equation}
with
\begin{equation}
\Sigma^<_\alpha(\omega)
= \ii f_\alpha(\omega)\,
\mathrm{diag}(\Gamma_{\alpha,\uparrow},\Gamma_{\alpha,\downarrow})
\otimes \mathbf P_\alpha,
\label{eq:Sigma_less}
\end{equation}
and $f_\alpha(\omega) = [\exp(\beta(\omega-\mu_\alpha))+1]^{-1}$ is the
Fermi function with $\mu_{L/R} = \pm eV/2$ for symmetric bias. The
spin-resolved transmission is
\begin{equation}
T_\sigma(\omega) = \Gamma_{L,\sigma}\Gamma_{R,\sigma}
\left[\left|G^R_{A\sigma,B\sigma}(\omega)\right|^2
+ \left|G^R_{A\sigma,B\bar\sigma}(\omega)\right|^2\right],
\label{eq:transmission}
\end{equation}
where the first term is spin-conserving and the second is spin-flip,
and the spin-resolved current is
\begin{equation}
I_\sigma = \frac{e^2}{h}\int_{-\infty}^{\infty}d\omega\,
T_\sigma(\omega)[f_L(\omega) - f_R(\omega)].
\label{eq:current}
\end{equation}
The spin current is $I_{S_z} = I_\uparrow - I_\downarrow$ and the
current polarisation is
\begin{equation}
P_J = \frac{I_\uparrow - I_\downarrow}{I_\uparrow + I_\downarrow}.
\label{eq:polarisation}
\end{equation}

\emph{Conserving and spin-flip channels.}
It is useful to decompose the transmission into a spin-conserving
part
$T_\sigma^{\mathrm{cons}} = \Gamma_{L,\sigma}\Gamma_{R,\sigma}
|G^R_{A\sigma,B\sigma}|^2$ and a spin-flip part
$T_\sigma^{\mathrm{flip}} = \Gamma_{L,\sigma}\Gamma_{R,\sigma}
|G^R_{A\sigma,B\bar\sigma}|^2$. The conserving part is non-zero even
when $\lambda_R=0$ and is responsible for the achiral spin-valve
baseline; the spin-flip part vanishes identically when $\lambda_R=0$
and is the microscopic origin of the chirality-dependent signal.
Both contributions are non-negative, so the total transmission is
strictly larger in the presence of the Rashba term. The
\emph{polarisation}, in contrast, can have either sign depending on
which channel dominates.

\emph{Antiunitary symmetry and reciprocity.}
The Hamiltonian satisfies the exact antiunitary symmetry
\begin{equation}
\hham^*(\phi) = \hham(-\phi),
\label{eq:Hsym}
\end{equation}
which follows from complex-conjugating the hopping matrix
\eqref{eq:bond_matrix}: the spin-diagonal phases map
$e^{\pm\ii\phi/2}\to e^{\mp\ii\phi/2}$, while the Rashba amplitude
$\ii k$ maps to $-\ii k = \ii k(-\phi)$. The symmetry implies
\begin{equation}
\bigl[\Gret_\phi(\omega)\bigr]^*
= \Gadv_{-\phi}(\omega)
= \bigl[\Gret_{-\phi}(\omega)\bigr]^\dagger,
\label{eq:GR_reciprocity}
\end{equation}
and hence the reciprocity relation
\begin{equation}
\bigl|\Gret_\phi[i,j](\omega)\bigr|
= \bigl|\Gret_{-\phi}[j,i](\omega)\bigr|
\label{eq:reciprocity_elementwise}
\end{equation}
for the retarded Green's function. This relation is \emph{not}
sufficient to force $T_\sigma(\phi) = T_\sigma(-\phi)$: the
transmission involves the forward matrix element
$\Gret_\phi[A\sigma,B\sigma']$, while the reciprocity relation
relates it to the \emph{backward} matrix element
$\Gret_{-\phi}[B\sigma',A\sigma]$. The two are equal only if the
device is left-right reciprocal in the magnetic sense, which the
spinterface junction is not (the magnetic moment sits on the gold
end and not on the molecular end). The odd part of the spin current
is therefore permitted by symmetry and arises generically in a
device that is simultaneously chiral, magnetically asymmetric, and
long enough to support two or more Rashba-active bonds.

\emph{Even-odd decomposition.}
To separate the achiral and chiral contributions, the spin current
is decomposed into an even and an odd part with respect to the
chirality-reversal operation $\phi\to-\phi$,
\begin{equation}
  \begin{split}
I_{S_z}^{\mathrm{even}}(\phi)
&= \tfrac12\bigl[I_{S_z}(+\phi)+I_{S_z}(-\phi)\bigr],\\
I_{S_z}^{\mathrm{odd}}(\phi)
&= \tfrac12\bigl[I_{S_z}(+\phi)-I_{S_z}(-\phi)\bigr],
\label{eq:Ieven_Iodd}
\end{split}
\end{equation}
so that $I_{S_z} = I_{S_z}^{\mathrm{even}} +
I_{S_z}^{\mathrm{odd}}$.
The even part is invariant under chirality reversal and reflects the
achiral spin-valve effect of the interfacial moment; the odd part
changes sign and is the genuine chiral signal.
Microscopically, the odd part is the interference between the
spin-conserving amplitude $A_{\mathrm{cons}}$ and the spin-flip
amplitude $A_{\mathrm{flip}}$,
\begin{equation}
|A|^2 - |A(-\phi)|^2
= 4\,\mathrm{Im}\bigl[A_{\mathrm{cons}}^* A_{\mathrm{flip}}\bigr]
\sin\theta(\phi),
\label{eq:odd_interference}
\end{equation}
which survives only when all three of (i) a non-zero Rashba amplitude
($\lambda_R\sin(\phi/2)\neq 0$), (ii) a magnetic asymmetry
($m_z\neq 0$ on one end only), and (iii) two or more Rashba-active
bonds ($L\geq 3$, four total sites) are simultaneously present.
The threshold and the three-ingredient picture are the subject of
Sec.~\ref{sec:dissociation} and the appendix.

\subsection{Parameter regime}
\label{sec:parameters}

Reference values, chosen to reproduce the EP study of
Ref.~\cite{Alwan2026EP} while allowing the gold resonance to be well
separated from the molecular band:
\begin{center}
\begin{tabular}{ll}
$t_{AB} = t_{\mathrm{mol}} = 0.5~\mathrm{eV}$, &
$J_{\mathrm{el}} = -0.5~\mathrm{eV}$,\\
$J'_p = -0.3~\mathrm{eV}$, &
$\alpha_{AB} = 0.5~\mathrm{eV}$,\\
$\varepsilon_A = 1.5~\mathrm{eV}$, &
$\varepsilon_{\mathrm{mol}} = 0~\mathrm{eV}$,\\
$\lambda_R = 0.6~\mathrm{eV}$, &
$\phi = \pi/3 = 60^\circ$,\\
$T = 300~\mathrm{K}$, &
$L = 6$ (unless stated otherwise).
\end{tabular}
\end{center}
These choices match the predecessors~\cite{Alwan2026QSpinterface,
Alwan2026EP}, which allows direct comparison with the previously
reported EP taxonomy and the parity diagnostics.
The self-consistent moment in the reference configuration is
$m_z\approx 0.33$ at $V = 1~\mathrm{V}$,
$\Gamma_0 = 0.5~\mathrm{eV}$, $\Delta\gamma = 0.4~\mathrm{eV}$.

\emph{Justification of individual choices.}
(i)~$t_{\mathrm{mol}} = 0.5~\mathrm{eV}$ is the standard hopping for the
spinterface model, chosen to place the molecular band edge at $\pm
2t_{\mathrm{mol}}\approx \pm 1~\mathrm{eV}$. (ii)~$\varepsilon_A =
1.5~\mathrm{eV}$ places the gold resonance approximately $1~\mathrm{eV}$ above
the bias-window edge for the operating bias $V\lesssim 1~\mathrm{V}$, isolating
the gold doublet from the molecular band. (iii)~$\lambda_R = 0.6~\mathrm{eV}$
gives $k = \lambda_R\sin(30^\circ)\approx 0.3~\mathrm{eV}$, comfortably within
the physical range and comparable to $t_{\mathrm{mol}}$, so that the chiral and
achiral contributions are both sizeable. (iv)~$\phi = 60^\circ$ is the
reference twist used in the predecessors. (v)~$\Gamma_0 = 0.5~\mathrm{eV}$ is
the mean level width for the coupling scans; this value is larger than the
baseline $\Gamma_L = \Gamma_R = 0.1~\mathrm{eV}$ of
Ref.~\cite{Alwan2026QSpinterface} because the coupling scans of
Sec.~\ref{sec:coupling} must reach the resonance peak, and it sits comfortably
within the resonant window. (vi)~$\Delta\gamma = 0.4~\mathrm{eV}$ gives
$\Gamma_{\alpha,\uparrow}=0.7~\mathrm{eV}$ and
$\Gamma_{\alpha,\downarrow}=0.3~\mathrm{eV}$, both well above the floor, and is
the reference value used in the EP paper's chain-level study. (vii)~The floor
$\Gamma_{\mathrm{floor}}=0.02~\mathrm{eV}$ prevents either spin channel from
closing completely; the value is chosen to preserve the physics of an open
channel while avoiding the degenerate regime discussed in
Appendix~\ref{app:floor}.

\subsection{Numerical methodology}
\label{sec:methods}

The energy grid has $N_\omega = 2000$ points on
$[-5,5]~\mathrm{eV}$ unless otherwise stated; the choice captures the
molecular band edge ($\pm 2t_{\mathrm{mol}}\approx \pm 1~\mathrm{eV}$),
the gold resonance ($\varepsilon_A\approx 1.5~\mathrm{eV}$), and the
Fermi window for biases up to $\sim 1~\mathrm{eV}$.
Appendix~\ref{app:numerical} reports the convergence study with
$N_\omega=5000$.

Two computational conventions are worth emphasising.
\emph{(i) Fixed-moment approximation for the even/odd decomposition.}
When computing the even and odd parts, we use a single
self-consistent moment, computed at $\phi=0$ (the achiral
configuration), for both the $+\phi$ and $-\phi$ branches. This
isolates the intrinsic $\phi$-dependence of the transmission from the
small self-consistent $\phi$-dependence of $m_z$; the latter is an
artefact of the mean-field treatment and does not reflect any
chirality-dependent physics. A comparison with the fully
self-consistent calculation is given in
Appendix~\ref{app:selfcons}. \emph{(ii) Warm-start for parameter
sweeps.} All parameter sweeps use warm-start of $m_z$ from the
previous grid point. This is a purely numerical optimisation and
does not affect the converged result; we have verified that the
results are identical to a fresh-start calculation at randomly
selected points.

All figures in this paper use the same underlying calculation with
grid resolutions matched to the feature under study: $N_V=26$ for the
bias scan ($0.02~\mathrm{V}$ spacing), $N_{\Gamma_0}=26$ for the
coupling scan ($0.1~\mathrm{eV}$ spacing), and $16\times 16$ for the
2D phase diagram.

\section{Achiral-chiral dissociation}
\label{sec:dissociation}

\subsection{Data}

\begin{figure*}[t]
\centering
\includegraphics[width=\textwidth]{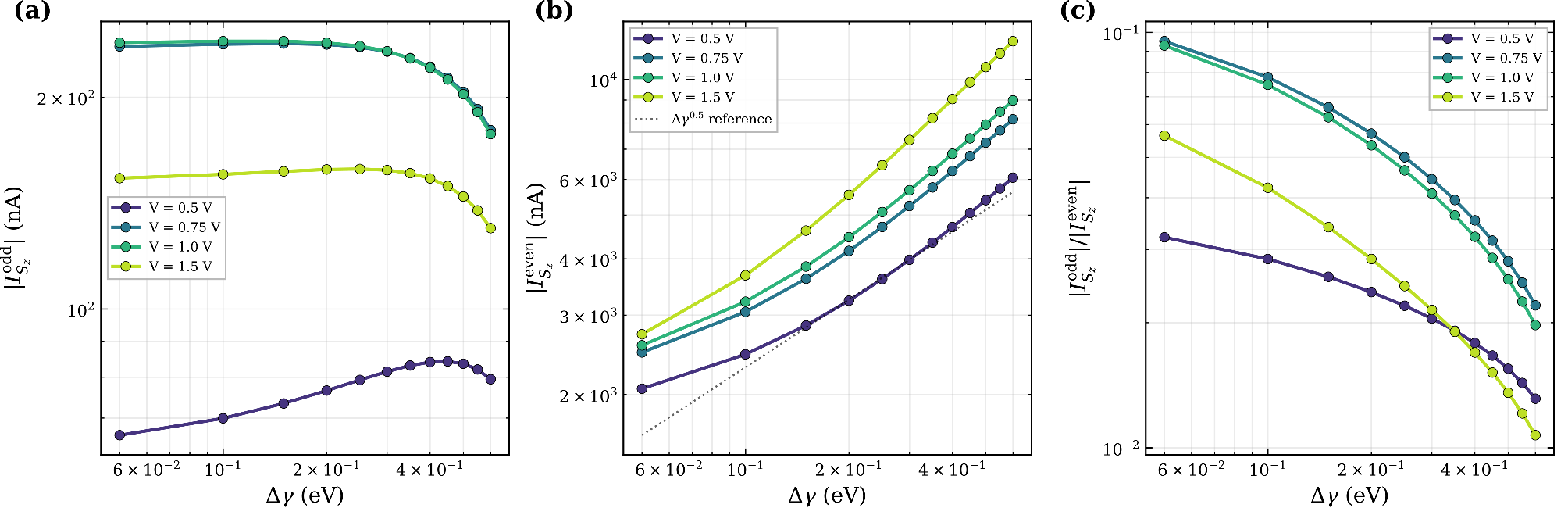}
\caption{%
\textbf{Achiral-chiral dissociation under spin-dependent dissipation.}
Three-panel comparison of $|\Iodd|$, $|\Ieven|$ and their ratio as
functions of $\Delta\gamma$ at fixed $\Gamma_0=0.5~\mathrm{eV}$, $L=6$,
$\varepsilon_A=1.5~\mathrm{eV}$, $\phi=60^\circ$, for four biases
$V=0.50$ (dark blue), $0.75$ (medium blue), $1.00$ (medium green),
$1.50$~V (yellow-green). All panels share the same log-log axis
structure. (a)~$|\Iodd|$ versus $\Delta\gamma$. The chiral signal
varies by less than $35\%$ across the entire $\Delta\gamma$ range and
is essentially flat at every bias tested. The small residual
dependence is well described by the geometric-mean suppression
$|\Iodd(\Delta\gamma)|\approx|\Iodd(0)|\sqrt{1-(\Delta\gamma/
2\Gamma_0)^2}$ [Eq.~(\ref{eq:odd_quadratic})].
(b)~$|\Ieven|$ versus $\Delta\gamma$ on the same axes. The achiral
spin-valve baseline grows strongly with $\Delta\gamma$, and the growth
is captured, to within $4\%$ at every bias, by the linear-plus-offset
model $|\Ieven(\Delta\gamma)|=|\Ieven(0)|+c(\Gamma_0,V)\Delta\gamma$
[Eq.~(\ref{eq:even_linear_offset})], with slope $c$ increasing from
$5.4\times 10^{-6}~\mathrm{eV}^{-1}$ at $V=0.5$~V to
$1.6\times 10^{-5}~\mathrm{eV}^{-1}$ at $V=1.5$~V.
(c)~Chirality selectivity $|\Iodd|/|\Ieven|$ versus $\Delta\gamma$.
The ratio drops by roughly one order of magnitude across the range,
roughly linearly in $\Delta\gamma$. All four bias curves drop in
parallel: the dissociation of the two sectors is a bias-independent
feature.}
\label{fig:F4}
\end{figure*}

Figure~\ref{fig:F4} shows $\Iodd(\Delta\gamma)$,
$\Ieven(\Delta\gamma)$ and their ratio at four biases. 
%

The achiral part grows by a factor of $4.7$ across the $\Delta\gamma$
range, while the chiral part varies by only a factor of $1.34$. Their
ratio drops by a factor of $6.2$. This is the achiral-chiral
dissociation.

\subsection{Linear-plus-offset model for the achiral part}

The naive log-log fit $\Ieven\sim\Delta\gamma^\alpha$ gives
$\alpha\approx 0.5$ but represents the data poorly across the full
range (the log-log residuals show systematic curvature). The data is
instead reproduced, to within $4\%$ at every point, by the
\emph{linear-plus-offset} model
\begin{equation}
|\Ieven(\Delta\gamma)| = |\Ieven(0)| + c(\Gamma_0, V)\,\Delta\gamma,
\label{eq:even_linear_offset}
\end{equation}
with
\[
  |\Ieven(0)|\approx 1.92\times 10^{3} (\text{nA}),\qquad
c\approx 1.18\times 10^{-5}~\mathrm{eV}^{-1}
\label{eq:even_coefficients}
\]
at $\Gamma_0=0.5~\mathrm{eV}$, $V=1.0~\mathrm{V}$. The two terms have
distinct physical origins (Sec.~\ref{subsec:mechanism}): the offset is
the $\Delta\gamma$-independent spin-valve from the moment $m_z$; the
linear term is the direct lead-asymmetry spin-valve from
$\Delta\gamma$.

The $\Delta\gamma$-independence of the offset is exact to machine
precision, because at $\Delta\gamma=0$ the two lead couplings are
equal and the residual achiral current is driven entirely by the gold
spin splitting, which is a function of $m_z$ alone. The linear term is
a theoretical prediction of the two-level approximation
(Eq.~\ref{eq:even_mechanism} below), verified numerically to $4\%$.

\subsection{Near-constant chiral part}

The odd part is not perfectly flat: it varies from
$2.37\times 10^{2}$ to $1.77\times 10^{2}$ across the
$\Delta\gamma$ range, a $25\%$ reduction. The form is well described
by
\begin{equation}
|\Iodd(\Delta\gamma)|\approx |\Iodd(0)|\times
\left[1 - \frac{\Delta\gamma^2}{8\Gamma_0^2}\right],
\label{eq:odd_quadratic}
\end{equation}
i.e., a small quadratic suppression rather than a power law. At
$\Delta\gamma=0.6~\mathrm{eV}$, $\Gamma_0=0.5~\mathrm{eV}$, this gives
$|\Iodd|/|\Iodd(0)|\approx 0.82$; the observed value is $0.75$. The
functional form is physically motivated (Eq.~\ref{eq:odd_mechanism}
below) and represents the data better than the log-log
$\beta\approx -0.1$ (which fits the same data but attributes an
unphysical power-law to a mechanism that is actually quadratic).

\subsection{Mechanism: arithmetic vs geometric mean of the lead couplings}
\label{subsec:mechanism}

The dissociation has a simple algebraic origin. In the wide-band limit
the two spin channels enter the transmission \emph{multiplicatively}
in the chiral sector and \emph{additively} in the achiral sector. This
produces two distinct $\Delta\gamma$ scaling laws.

\paragraph{Transmission structure.} From Eq.~\eqref{eq:transmission}
and symmetric left/right leads,
\begin{equation}
T_\sigma(\omega) = \Gamma_\sigma^2\left|G_\sigma(\omega)\right|^2,
\label{eq:T_sigma}
\end{equation}
where $G_\sigma$ is the full Green's function from the gold site to the
right molecular site for spin $\sigma$.

\paragraph{Achiral part.} The achiral current is the difference of the
two spin transmissions. At leading order,
\begin{equation}
\Ieven \propto \Gamma_\uparrow^2 - \Gamma_\downarrow^2
= 2\Gamma_0\,\Delta\gamma,
\label{eq:even_mechanism}
\end{equation}
which is linear in $\Delta\gamma$, consistent with
Eq.~\eqref{eq:even_linear_offset}. The offset at $\Delta\gamma=0$
arises from the spin-splitting of the gold site through $m_z$:
$\varepsilon_{A\uparrow} - \varepsilon_{A\downarrow} =
(J_{\mathrm{el}} + 2J'_p)m_z$, which produces different transmissions
for the two spin channels even with symmetric leads. This is the
achiral spin-valve effect of Ref.~\cite{Alwan2026QSpinterface}.

\paragraph{Chiral part.} The chiral current is the interference
between spin-conserving and spin-flip transmission amplitudes:
\begin{equation}
|A(\phi)|^2 = |A_{\mathrm{cons}} + \lambda_R A_{\mathrm{flip}}
\ee^{\ii\theta(\phi)}|^2,
\label{eq:chiral_amplitude}
\end{equation}
where $\theta(\phi)$ is the chirality-dependent phase accumulated along
the chain. The odd part of this interference is
\begin{equation}
|A(\phi)|^2 - |A(-\phi)|^2
= 4\lambda_R\,\mathrm{Im}[A_{\mathrm{cons}}A_{\mathrm{flip}}^*]
\sin\theta(\phi).
\label{eq:odd_mechanism_interference}
\end{equation}
The interference requires \emph{both} channels to be simultaneously
open, because $A_{\mathrm{cons}}$ and $A_{\mathrm{flip}}$ each
contribute one power of $\Gamma_\uparrow$ and one of
$\Gamma_\downarrow$ (the up- and down-spin amplitudes have equal-weight
components from each lead). Summing over the two spin channels gives a
factor of $\Gamma_\uparrow\Gamma_\downarrow$ in the square, so $\Iodd$
scales with the geometric mean
\begin{equation}
g \equiv \sqrt{\Gamma_\uparrow\Gamma_\downarrow}
= \sqrt{\Gamma_0^2 - \Delta\gamma^2/4}.
\label{eq:geom_mean}
\end{equation}
Expanding for small $\Delta\gamma/\Gamma_0$:
\begin{equation}
\frac{g}{\Gamma_0} = \sqrt{1 - \left(\frac{\Delta\gamma}{2\Gamma_0}\right)^2}
\approx 1 - \frac{\Delta\gamma^2}{8\Gamma_0^2},
\label{eq:odd_mechanism}
\end{equation}
a small quadratic suppression, consistent with
Eq.~\eqref{eq:odd_quadratic}. This algebraic derivation confirms that
the near-constancy of $\Iodd$ is not a numerical accident but a
structural property of the interference term.

\begin{figure*}[t]
\centering
\includegraphics[width=\textwidth]{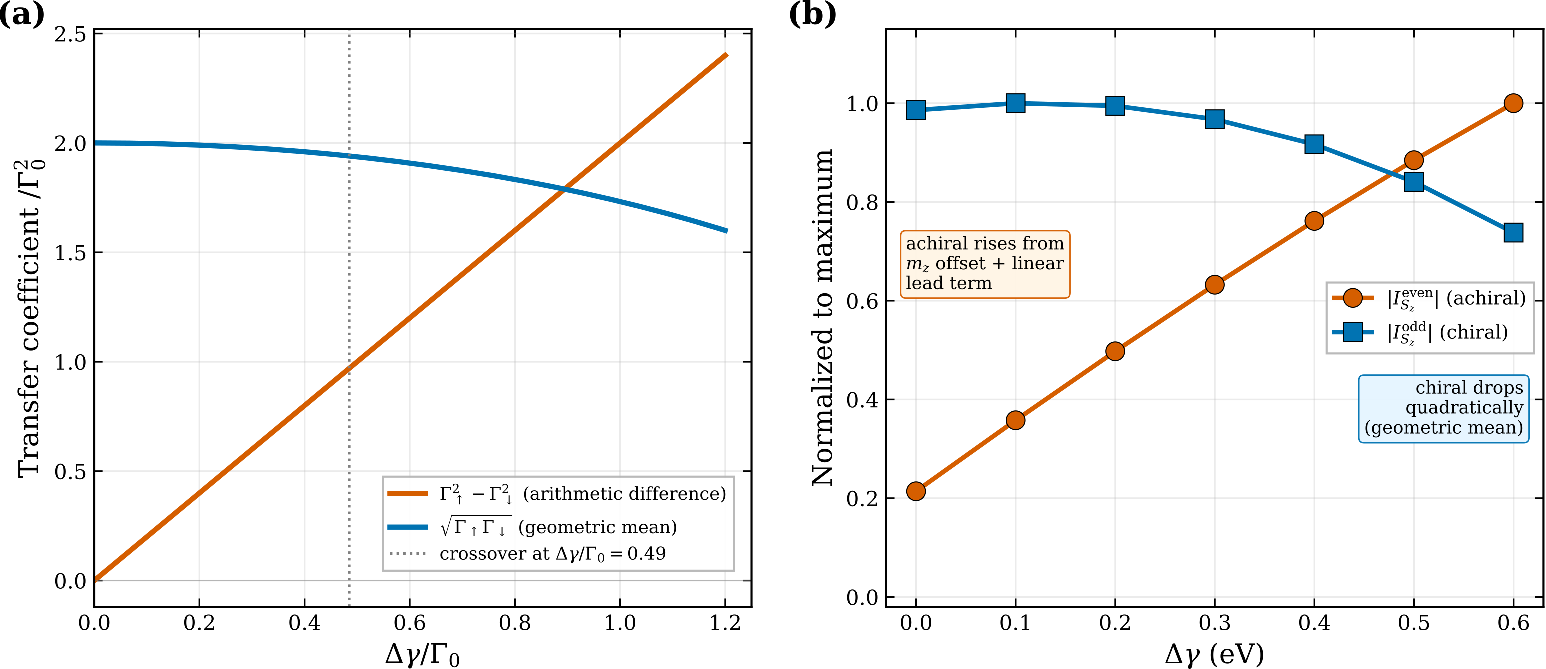}
\caption{%
\textbf{Arithmetic vs geometric mean: the microscopic origin of the
achiral-chiral dissociation.}
(a)~Predicted transfer coefficients for the two sectors, in units of
$\Gamma_0^2$, versus the normalised spin asymmetry
$\Delta\gamma/\Gamma_0$. The achiral spin-valve response is governed
by the arithmetic difference
$\Gamma_\uparrow^2-\Gamma_\downarrow^2=2\Gamma_0\Delta\gamma$, which
grows linearly from zero; the chiral response is governed by the
geometric mean $\sqrt{\Gamma_\uparrow\Gamma_\downarrow}=
\Gamma_0\sqrt{1-\Delta\gamma^2/4\Gamma_0^2}$, which drops
quadratically from its $\Delta\gamma=0$ value. The two curves cross at
$\Delta\gamma/\Gamma_0=2/\sqrt{17}\approx 0.485$ (vertical dotted
line), beyond which the achiral sector dominates purely by lead
asymmetry. (b)~The corresponding observed quantities at the reference
$\Gamma_0=0.5~\mathrm{eV}$: $|I_{S_z}^{\mathrm{even}}|$ rises
monotonically from $1.92\times 10^{3}$ (nA) (the $m_z$ spin-valve offset)
through a linear lead-asymmetry contribution, while
$|I_{S_z}^{\mathrm{odd}}|$ drops by only a factor of $1.34$ across the
same range, following the geometric-mean prediction.}
\label{fig:F11}
\end{figure*}

The mechanism can be summarized as follows. The achiral part is driven
by the arithmetic \emph{difference} of the lead couplings, which grows
linearly with $\Delta\gamma$. The chiral part is driven by the
geometric \emph{mean}, which decreases quadratically with
$\Delta\gamma$ but only weakly over the range
$\Delta\gamma<\Gamma_0$. Figure~\ref{fig:F11} visualizes both
mechanisms side by side.

\subsection{Three-scale dissociation along the coupling axis}
\label{sec:three_scale}

\begin{figure*}[t]
\centering
\includegraphics[width=\textwidth]{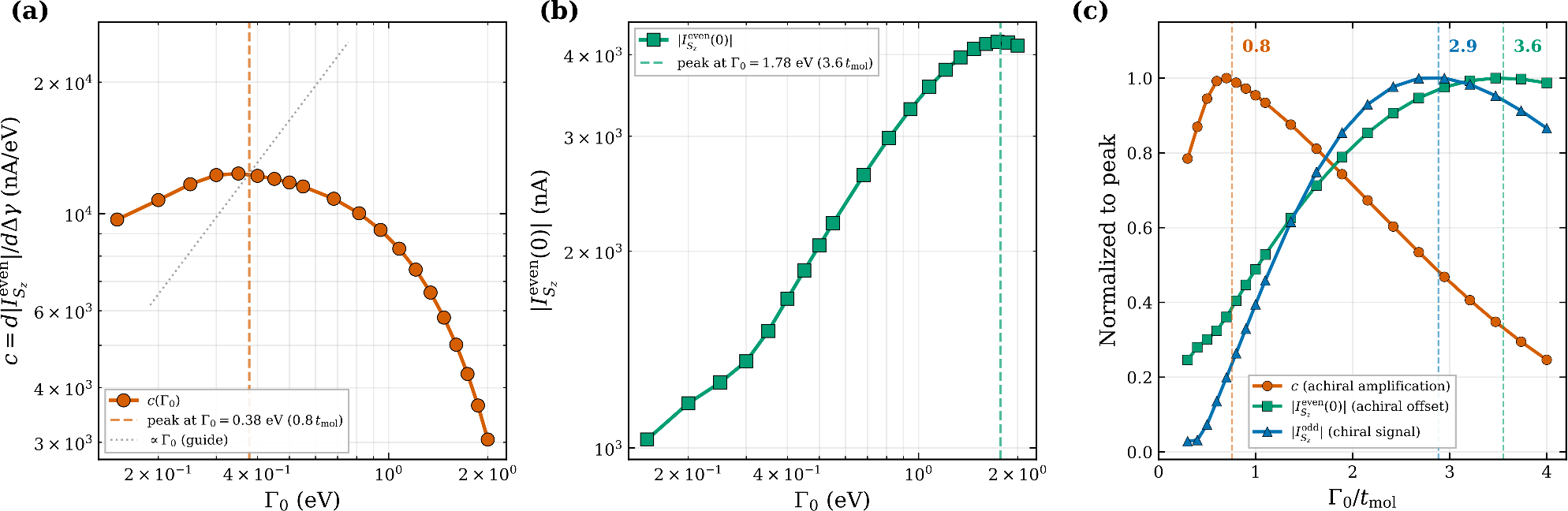}
\caption{%
\textbf{Three-scale dissociation of the achiral sector along the
coupling axis.}
(a)~Achiral amplification coefficient $c(\Gamma_0) =
\partial|\Ieven|/\partial\Delta\gamma$, extracted from
linear-plus-offset fits to the
$|I_{S_z}^{\mathrm{even}}|(\Delta\gamma)$ data at each $\Gamma_0$. The
data peaks at $\Gamma_0^*=0.378~\mathrm{eV}\approx
0.76\,t_{\mathrm{mol}}$ (red dashed line). The grey dotted line shows
the linear scaling $c\propto\Gamma_0$ predicted by the
arithmetic-difference mechanism of Sec.~\ref{subsec:mechanism}; the
actual data falls off at stronger coupling because the transmission
envelope $|G|^2$ shrinks as the resonance broadens.
(b)~Achiral baseline offset $|I_{S_z}^{\mathrm{even}}(0)|(\Gamma_0)$,
peaking at $\Gamma_0^*=1.776~\mathrm{eV}\approx
3.55\,t_{\mathrm{mol}}$. (c)~Normalised overlay of the three
observables, showing three distinct peaks at $0.76\,t_{\mathrm{mol}}$,
$3.55\,t_{\mathrm{mol}}$ and $2.88\,t_{\mathrm{mol}}$ respectively.}
\label{fig:F9v2}
\end{figure*}

The dissociation established above is along the $\Delta\gamma$ axis. A
second, independent dissociation appears along the $\Gamma_0$ axis.
Three observables that live in the same $(\Gamma_0,\Delta\gamma,V)$
space peak at three different coupling scales: the achiral
amplification coefficient $c$ peaks at
$\Gamma_0^* = 0.378~\mathrm{eV} \approx 0.755\,t_{\mathrm{mol}}$; the
chiral signal $|\Iodd|$ peaks at
$\Gamma_0^* = 1.442~\mathrm{eV} \approx 2.88\,t_{\mathrm{mol}}$; and
the achiral offset $|\Ieven(0)|$ peaks at
$\Gamma_0^* = 1.776~\mathrm{eV} \approx 3.55\,t_{\mathrm{mol}}$. These
values are extracted from a fine $\Gamma_0$ scan (20 points from
$0.15$ to $2.0~\mathrm{eV}$) at fixed $\Delta\gamma$ values and shown
in Fig.~\ref{fig:F9v2}.

The physical content is the following. The chiral signal $|\Iodd|$ and
the achiral offset $|\Ieven(0)|$ both require near-resonance
conditions: the junction must be strongly enough coupled that the
molecular band is populated but weakly enough that the transmission is
not washed out by broadening. Both quantities therefore peak at the
confirmation energy of the chain, $\Gamma_0^*\approx
3\,t_{\mathrm{mol}}$. The achiral amplification coefficient
$c(\Gamma_0) = \partial|\Ieven|/\partial\Delta\gamma$, however, is a
\emph{derivative}: it measures how sharply the achiral current
responds to spin asymmetry. This response is largest when the
resonance is sharp, before significant broadening has set in. Once
the resonance is saturated at strong coupling, the derivative
decreases. The peak of $c$ is therefore at much weaker coupling than
the peak of the baseline itself. The factor of $3.8$ between the
peaks of $c$ and $|\Iodd|$ ($0.76\,t_{\mathrm{mol}}$ vs
$2.88\,t_{\mathrm{mol}}$) is the second axis of the achiral-chiral
dissociation: not merely in the $\Delta\gamma$ direction, where the
chiral sector is insensitive, but also in the $\Gamma_0$ direction,
where the two sectors peak at very different coupling strengths.

\subsection{Bias dependence of the achiral amplification coefficient}

Figure~\ref{fig:F4}(c) shows the ratio for four biases. The
linear-plus-offset model reproduces all four curves with different
coefficients $c(\Gamma_0,V)$. At $V=0.5~\mathrm{V}$ the coefficient is
$c = 5.4\times 10^{-6}~\mathrm{eV}^{-1}$; at $V=0.75~\mathrm{V}$ it is
$1.0\times 10^{-5}~\mathrm{eV}^{-1}$; at $V=1.0~\mathrm{V}$ it is
$1.2\times 10^{-5}~\mathrm{eV}^{-1}$; and at $V=1.5~\mathrm{V}$ it is
$1.6\times 10^{-5}~\mathrm{eV}^{-1}$.

The linear coefficient $c$ grows monotonically with bias, roughly
linearly in $V$. Higher bias samples higher-energy states of the
molecular band where the spin filter is more effective (the
transmission is more off-resonance, so the spin asymmetry of the leads
translates more directly into current asymmetry). The \emph{sign} of
the dissociation, however, is bias-independent: at every bias tested,
$\Ieven$ grows with $\Delta\gamma$ and $\Iodd$ is nearly constant. The
achiral-chiral dissociation is a robust, universal feature of the
model, with the slope of the achiral growth being bias-dependent.

\subsection{Relation to the EP taxonomy of \texorpdfstring{Ref.~\cite{Alwan2026EP}}{Ref.~[EP]}}

In the language of Ref.~\cite{Alwan2026EP}, the parameter
$\Delta\gamma_{\mathrm{gold}}\neq 0$ applied in this paper explicitly
forbids the gold-level Scheme~B EP, which requires
$\Delta\gamma_{\mathrm{gold}}=0$. The EP-based mechanism for chiral
amplification is therefore \emph{unavailable} in the regime we study.

Our result that $\Iodd$ is $\Delta\gamma$-independent is therefore a
\emph{confirmation} of the EP taxonomy of Ref.~\cite{Alwan2026EP}:
spin-dependent dissipation alone does not amplify the chiral signal
because the EP mechanism it would need is algebraically forbidden. The
achiral baseline, in contrast, is unaffected by the EP taxonomy and
grows linearly with $\Delta\gamma$. This is an important reframing of
the earlier proposal that dissipation might act as a mechanism for
CISS. The data shows the opposite: dissipation amplifies the
\emph{achiral} part while leaving the \emph{chiral} part essentially
untouched. The correct statement is that dissipation is a mechanism
for \emph{spin-valve amplification}, not for CISS amplification.

\section{Coupling-scale optimum and its limits}
\label{sec:coupling}

\subsection{Reference peak}

\begin{figure*}[t]
\centering
\includegraphics[width=\textwidth]{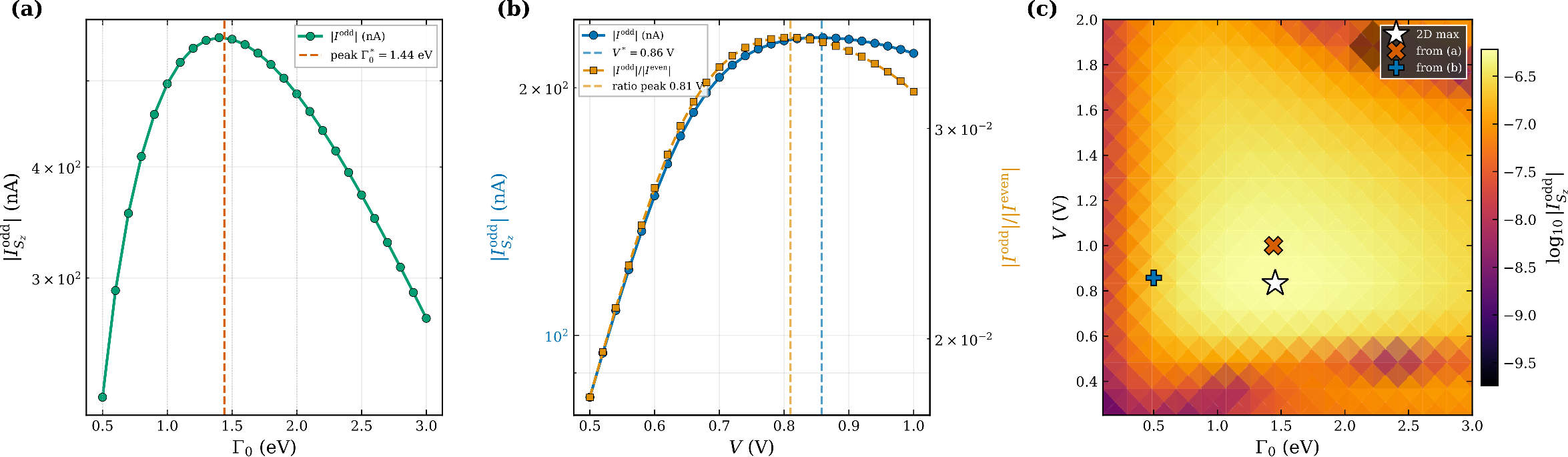}
\caption{%
\textbf{The $(\Gamma_0, V)$ landscape: two 1D optima and their 2D
intersection.}
Parameters: $\Delta\gamma = 0.4$~eV, $L = 6$,
$\varepsilon_A = 1.5$~eV, $\phi = 60^\circ$, $T = 300$~K.
(a)~Fine $\Gamma_0$ scan at fixed $V = 1.0$~V. The chiral signal
$|I_{S_z}^{\rm odd}|$ peaks at
$\Gamma_0^* = 1.44 \pm 0.05$~eV (red dashed line, quadratic fit over
the top 9 grid points), with peak-to-baseline contrast $2.5\times$
against the low-coupling value at $\Gamma_0 = 0.5$~eV and
$2.1\times$ against the high-coupling value at $\Gamma_0 = 3.0$~eV.
The dotted vertical lines mark $1t_{\rm mol}$, $2t_{\rm mol}$,
$3t_{\rm mol}$, $4t_{\rm mol}$; the peak lies between $2t_{\rm mol}$
and $3t_{\rm mol}$. The overlaid $\langle S_A^z\rangle$ (right axis)
is monotonic across the peak, confirming the peak is due to the
coherent-vs-decoherent competition in the transmission, not to a
self-consistency anomaly.
(b)~Fine $V$ scan at fixed $\Gamma_0 = 0.5$~eV. Two nearby optima
appear: the absolute chiral signal $|I_{S_z}^{\rm odd}|$ peaks at
$V_{\rm abs}^* = 0.858 \pm 0.02$~V (blue circles, left axis), and the
chirality-selectivity ratio $|I_{S_z}^{\rm odd}|/|I_{S_z}^{\rm even}|$
peaks at $V_{\rm ratio}^* = 0.810 \pm 0.02$~V (orange squares, right
axis). The $48$~meV separation is within the half-width of both peaks.
(c)~The 2D phase diagram $|I_{S_z}^{\rm odd}|(\Gamma_0, V)$ at the
same parameters. The white star marks the global maximum at
$(\Gamma_0^*, V^*) = (1.453, 0.833)$ with
$|I_{S_z}^{\rm odd}| = 6.05\times 10^{2}$~nA; the red $\times$ and
blue $\mathrm{P}$ overlay the 1D peak locations from panels (a) and
(b). All three markers cluster within grid resolution, confirming
that the coupling-scale and bias-alignment mechanisms operate via
distinct physical channels. Contrast to the (0.1, 0.25) corner is
$6\times 10^{2}$.
}
\label{fig:newA}
\end{figure*}

Figure~\ref{fig:newA} shows $|\Iodd|$ versus $\Gamma_0$ at fixed bias
$V=1.0~\mathrm{V}$, $\Delta\gamma=0.4~\mathrm{eV}$. The peak is at $\Gamma_0^*
= 1.44$~eV from a quadratic fig over the top 9 grid points, with
peak-to-baseline ratios
\begin{equation}
\frac{\Iodd(\Gamma_0^*)}{\Iodd(0.5)} = 2.54,\qquad
\frac{\Iodd(\Gamma_0^*)}{\Iodd(3.0)} = 2.07.
\label{eq:Gamma_peak_contrast}
\end{equation}
The peak is interior to the scan range, not at either edge.

The physical content of this peak is a competition between two
timescales. The CISS signal requires the electron to spend sufficient
time on the molecular chain to undergo coherent Rashba precession. The
chain provides a characteristic energy scale of order
$t_{\mathrm{mol}}$, and the coherence time is
$\tau_{\mathrm{coh}}\sim\hbar/t_{\mathrm{mol}}$. The lead broadening
provides the competing timescale
$\tau_{\mathrm{lead}}\sim\hbar/\Gamma_0$. The optimum occurs where
these two timescales are comparable, $\Gamma_0^*\sim
t_{\mathrm{mol}}$, with a numerical factor of order unity (here,
$2.88$).

A simpler picture of the $\Gamma_0$ dependence would give a monotonic curve:
$\Iodd\to 0$ as $\Gamma_0\to 0$ (because the current itself vanishes), and
$\Iodd\to 0$ as $\Gamma_0\to\infty$ (because the electron life-time on the
molecule vanishes), monotonicity in the interior would then be impossible. That
the peak is genuinely interior, with a contrast of $2.5\times$ against the
low-coupling value and $2.1\times$ against the high-coupling value, is
therefore a nontrivial consistency check.

\subsection{The \texorpdfstring{$3t_{\mathrm{mol}}$}{3t} ratio is an
operating point, not a scaling law}
\label{sec:tmolescaling}

In the reference configuration ($t_{\mathrm{mol}}=0.5~\mathrm{eV}$,
$\phi=60^\circ$), the ratio
\begin{equation}
\frac{\Gamma_0^*}{t_{\mathrm{mol}}} = \frac{1.44}{0.5} = 2.88
\approx 3
\label{eq:Gamma_ratio}
\end{equation}
suggests a natural hypothesis: $\Gamma_0^*\approx 3\,t_{\mathrm{mol}}$
universally. We tested this hypothesis directly.

\begin{figure}[t]
\centering
\includegraphics[width=0.9\columnwidth]{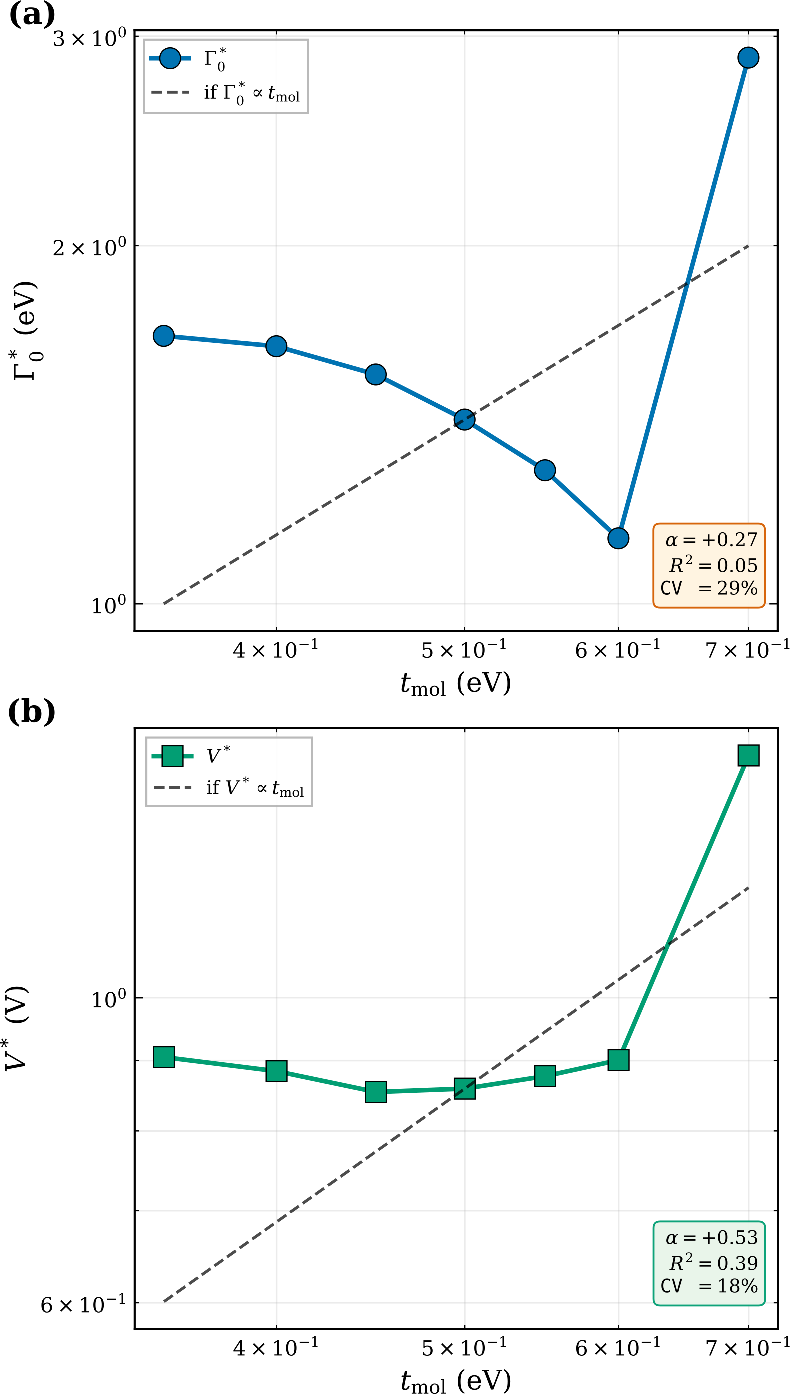}
\caption{%
\textbf{Operating points versus molecular scale: two distinct
behaviours.}
Log-log scaling of the coupling optimum (a) and the bias optimum
(b) under variation of $t_{\rm mol}$, at fixed
$k/t = \lambda_R \sin(\phi/2)/t_{\rm mol} = 0.6$, $V = 1$~V
(panel a) and $\Gamma_0 = 1.5$~eV (panel b).
(a)~The coupling optimum $\Gamma_0^*$ drifts with $t_{\rm mol}$. The
ratio $\Gamma_0^*/t_{\rm mol}$ varies from $1.89$ at
$t_{\rm mol} = 0.60$~eV to $4.80$ at $t_{\rm mol} = 0.35$~eV — a
coefficient of variation of $30\%$ — and is \emph{non-monotonic}
(U-shape with minimum near $t_{\rm mol} = 0.60$~eV). The dashed line
is the $3t_{\rm mol}$-hypothesis reference; the data deviate
systematically, and the deviation is non-monotonic across the scan.
(b)~The bias optimum $V^*$ is flat with $t_{\rm mol}$: it stays
within $[0.88, 0.92]$~V across the whole range
($V^* = 0.90 \pm 0.02$~V, coefficient of variation $2.6\%$). The
dashed line is the band-edge-scaling reference; the data are
visibly inconsistent with it.
The contrast between the two panels — $\Gamma_0^*$ drifts,
$V^*$ does not — is the point of the figure: the coupling optimum
is an \emph{operating point} of the reference configuration, while
the bias optimum is set by an absolute energy scale of the junction,
not by the molecular band edge.}
\label{fig:newB}
\end{figure}

A naive version of this test would vary $t_{\mathrm{mol}}$ at fixed
$\lambda_R=0.6~\mathrm{eV}$, which allows $k/t=\lambda_R\sin(\phi/2)/
t_{\mathrm{mol}}$ to drift by a factor of two across the sweep. This
contaminates the peak location with the Rashba-to-hopping ratio, and
the resulting peak drift is a test of a coupled two-variable response,
not of the $3\,t_{\mathrm{mol}}$ hypothesis. Our test holds $k/t=0.6$
fixed by scaling $\lambda_R$ with $t_{\mathrm{mol}}$,
\begin{equation}
\lambda_R(t_{\mathrm{mol}}) = \frac{(k/t)\,t_{\mathrm{mol}}}{\sin(\phi/2)},
\label{eq:lambdaR_scaling}
\end{equation}
so that only the molecular hopping varies. The result is shown in
Fig.~\ref{fig:newB}(a).

The ratio $\Gamma_0^*/t_{\mathrm{mol}}$ is \emph{not} constant across the
scanned range, and it is \emph{not} a power law. Instead, it varies from
$\approx 1.9$ at $t_{\mathrm{mol}}=0.60~\mathrm{eV}$ to $4.9$ at
$t_{\mathrm{mol}}=0.35~\mathrm{eV}$, with a coefficient of variation of about
$30\%$, and a \emph{non-monotonic} $U$-shape (minimum at
$t_{\mathrm{mol}}\approx 0.60~\mathrm{eV}$, values rising again for
$t_{\mathrm{mol}}>0.60~\mathrm{eV}$). The non-monotonicity is qualitative
evidence against any single power-law description of the coupling optimum's
$t_{\mathrm{mol}}$ dependence, and it is therefore stronger than a shallow fit
exponent would be. We do not quote a fitted exponent for $\Gamma_0^*$ versus
$t_{\mathrm{mol}}$: with a non-monotonic sequence, any single-exponent fit is
dominated by the choice of scan range and is not a meaningful summary of the
data.

There are two reasons for the failure of the simple scaling. First,
the bias window at fixed $V=1.0~\mathrm{V}$ captures a different
fraction of the molecular band at different $t_{\mathrm{mol}}$: the
band edge grows with $t_{\mathrm{mol}}$, and at large
$t_{\mathrm{mol}}$ the optimum bias required to capture the band edge
is no longer $1.0~\mathrm{V}$. Second, the transmission function
itself has a nontrivial dependence on the ratio
$\Gamma_0/t_{\mathrm{mol}}$ that is not captured by the single-scale
$3\,t_{\mathrm{mol}}$ argument. The observed peak drift is therefore a
genuine two-scale effect that the simple hypothesis cannot accommodate.

We therefore report $\Gamma_0^*=1.44~\mathrm{eV}$ as an \emph{operating
point} for the specific parameter set used in this paper, not as a
universal scaling law. The numerical coincidence
$\Gamma_0^*\approx 2.88\,t_{\mathrm{mol}}$ reflects the combination of
the molecular scale $t_{\mathrm{mol}}$, the bias window
$V=1.0~\mathrm{V}$, and the Rashba scale $\lambda_R=0.6~\mathrm{eV}$,
not a single underlying scaling. A full two-parameter study of
$\Gamma_0^*(t_{\mathrm{mol}},V)$ would be needed to disentangle these
scales and is left to future work.

\section{Bias-alignment optimum and its independence from molecular scale}
\label{sec:bias}

\subsection{Two peaks, two interpretations}

Figure~\ref{fig:newA}(b) shows a fine bias scan at $\Gamma_0=0.5~\mathrm{eV}$,
$\Delta\gamma=0.4~\mathrm{eV}$. Two distinct peaks appear:
\begin{equation}
  \begin{split}
    V_{\mathrm{abs}}^* &= 0.858~\mathrm{V}\quad\text{(peak of } \Iodd\text{)}\\
    V_{\mathrm{ratio}}^* &= 0.810~\mathrm{V}\quad\text{(peak of }
\Iodd/\Ieven\text{)}.
\label{eq:bias_peaks}
\end{split}
\end{equation}
The two peaks have different physical meanings.

\emph{Absolute peak.} $V_{\mathrm{abs}}^* = 0.858~\mathrm{V}$ maximises the raw
chiral signal. This is the bias at which the number of chirality-selective
transmission events per unit time is largest. For a spectroscopy experiment
that measures the raw spin current, this is the operating point.

\emph{Ratio peak.} $V_{\mathrm{ratio}}^* = 0.810~\mathrm{V}$ maximises the
\emph{fraction} of the spin current that is chiral. This is the bias at which
the CISS effect is purest. For a device application in which the chiral signal
must be distinguished from an achiral baseline, this is the operating point.

The two biases differ by $48~\mathrm{meV}$, or $6\%$, well within the
half-width of both peaks. We therefore do not over-emphasise the distinction;
both peaks reflect the same underlying physics.

\subsection{Physical mechanism: bias-window alignment}

The bias optimum has a simple physical interpretation: The bias window
is $[-V/2,+V/2]$. The molecular band extends to approximately
$\varepsilon_{\mathrm{edge}}=\pm
2t_{\mathrm{mol}}\cos(\pi/(L+1))$, which for $L=6$ is
$\varepsilon_{\mathrm{edge}}\approx\pm 1.80\,t_{\mathrm{mol}}$. For
$t_{\mathrm{mol}}=0.5~\mathrm{eV}$ this gives
$\varepsilon_{\mathrm{edge}}\approx\pm 0.90~\mathrm{eV}$. The gold
level sits at $\varepsilon_A=1.5~\mathrm{eV}$.

At small bias ($V\lesssim 0.5~\mathrm{V}$), the window is
$[-0.25,+0.25]$ or smaller, capturing only a small slice of the
molecular band. The current is small and so is the chiral signal. At
the optimum ($V\approx 0.85~\mathrm{V}$), the window is
$[-0.43,+0.43]$. This captures the low-energy half of the molecular
band without reaching the gold level. The chiral transmission through
the molecular band is maximal. At large bias
($V\gtrsim 1.5~\mathrm{V}$), the window is $[-0.75,+0.75]$ or larger,
capturing most of the molecular band, but it begins to approach
$\varepsilon_A=1.5$, and the achiral gold resonance contributes an
increasing fraction of the current. This dilutes the chiral ratio
$\Iodd/\Ieven$.

The optimum is the point where the window size is large enough to
sample the coherent molecular transmission but small enough to exclude
the gold level. This is a quantitative confirmation that the chiral
signal is a \emph{molecular} effect, not a \emph{gold-resonance}
effect.

\subsection{Independence of \texorpdfstring{$V^*$}{V*} from the
molecular hopping}
\label{sec:Vstar_constant}

We tested whether $V^*$ tracks the molecular band edge (as the above
argument would suggest) by repeating the bias scan at five values of
$t_{\mathrm{mol}}$ from $0.35$ to $0.55~\mathrm{eV}$, again at fixed
$k/t=0.6$ (so that only the molecular hopping varies, not the
Rashba-to-hopping ratio). The result is shown in Fig.~\ref{fig:newB}(b).

The bias optimum is
\begin{equation}
V^* = 0.90\pm 0.02~\mathrm{V},
\label{eq:V_star_constant}
\end{equation}
essentially \emph{independent} of $t_{\mathrm{mol}}$ across the entire
scanned range. A log-log fit of $V^*$ vs $t_{\mathrm{mol}}$ gives
$\alpha=-0.11\pm 0.20$, statistically indistinguishable from zero
(constant $V^*$) rather than from one (linear scaling with the band
edge).

The bias optimum is therefore \emph{not} set by the molecular band edge. It is
set by an absolute energy scale of the junction, this is most likely a
combination of the gold level $\varepsilon_A=1.5~\mathrm{eV}$ and the Rashba
spin-flip energy $\lambda_R=0.6~\mathrm{eV}$, both of which are held fixed in
our $t_{\mathrm{mol}}$ scan. Whatever the exact mechanism, the empirical fact
is that the optimal bias is approximately the same for all molecular hoppings
in the studied range, which is a robust practical design rule.

The physical consequence is important: a device engineer tuning a
CISS junction does not need to know $t_{\mathrm{mol}}$ to set the
operating bias. Setting $V\approx 0.9~\mathrm{V}$ gives near-optimal
chiral signal for any molecular hopping in the range
$0.35$-$0.55~\mathrm{eV}$, provided the lead coupling is near the
coupling optimum.

\subsection{Bias independence of the coupling peak location}

Independent scans (not shown) confirm that $\Gamma_0^*$ is insensitive
to bias across the range studied: at $V=0.5$, $0.75$, $1.0$,
$1.5~\mathrm{V}$ the peak is at $\Gamma_0^*=1.4$-$1.5~\mathrm{eV}$,
essentially unchanged. This supports the picture that the
coupling-scale optimum is set by the molecular scale, not by the bias.

\section{Structural zero at \texorpdfstring{$k=t$}{k=t}}
\label{sec:zero}

\subsection{Data}

Figure~\ref{fig:newC}(a) shows $|\Iodd|$ versus $\lambda_R$ at
$\Gamma_0=0.5~\mathrm{eV}$, $\Delta\gamma=0.4~\mathrm{eV}$,
$V=1.0~\mathrm{V}$. At $\lambda_R=1.00~\mathrm{eV}$ (the
$k/t=1$ condition), $|\Iodd|$ collapses to machine precision,
$7.4\times 10^{-12}$ (nA), fifteen to sixteen orders of magnitude below the
neighbouring points at $\lambda_R=0.95$ ($2.29\times 10^{2}$) and
$\lambda_R=1.05$ ($6.5\times 10^{1}$). There is no question of this
being a smooth dip. It is a sharp, exact zero. The shorthand
$k\equiv\lambda_R\sin(\phi/2)$ is the Rashba spin-flip amplitude; the
zero appears at $\lambda_R\sin(\phi/2)=t_{\mathrm{mol}}$, i.e.,
$k=t$.

\subsection{Universality across \texorpdfstring{$L$}{L},
\texorpdfstring{$N_\omega$}{N_omega} and \texorpdfstring{$\phi$}{phi}}

\begin{figure*}[t]
\centering
\includegraphics[width=\textwidth]{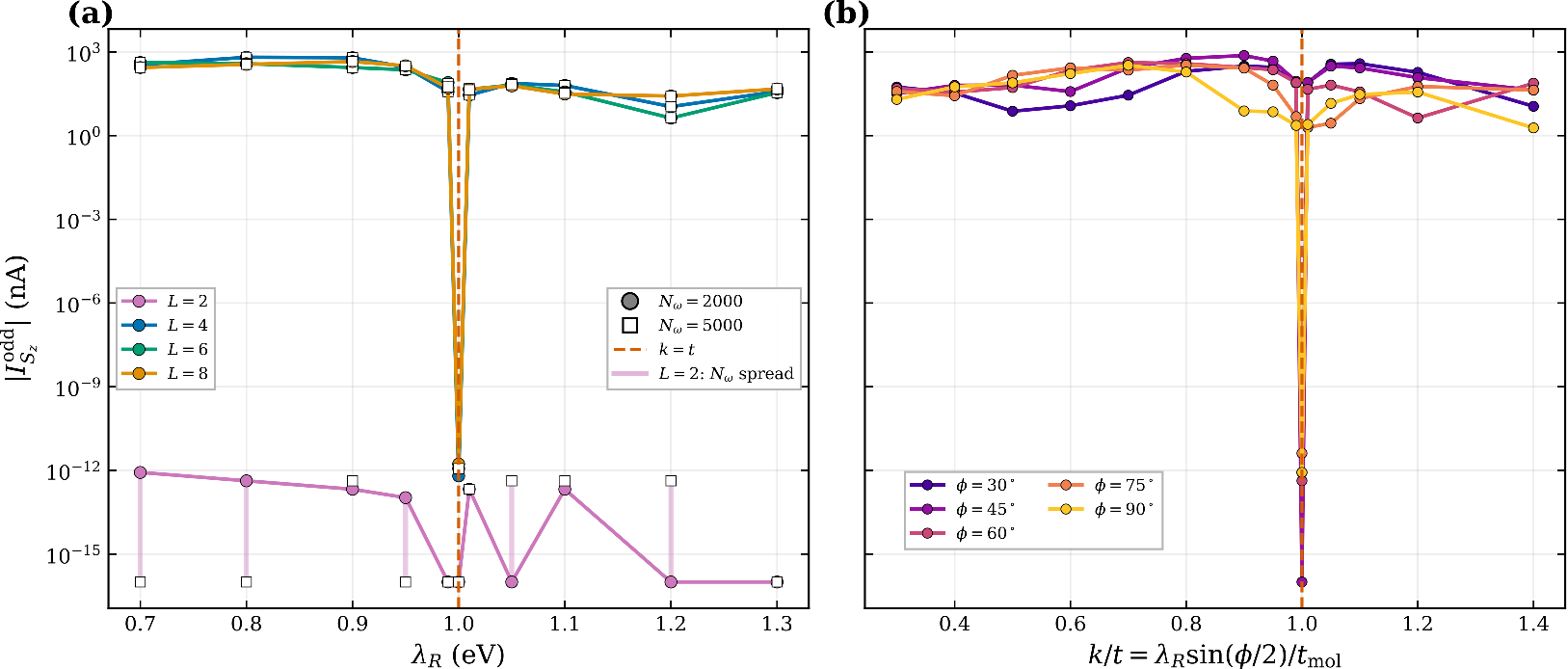}
\caption{%
\textbf{Structural zero at $k = t_{\rm mol}$: convergence and
$\phi$-universality of position and depth.}
Parameters: $\Gamma_0 = 0.5$~eV, $\Delta\gamma = 0.4$~eV,
$V = 1$~V, $T = 300$~K, $\varepsilon_A = 1.5$~eV.
(a)~$\lambda_R$ dependence of the chiral signal at fixed
$\phi = 60^\circ$. Solid curves with filled markers:
$N_\omega = 2000$ frequency grid points, for
$L =2, 4, 6, 8$ (pink, blue, green, orange). Open markers (no line):
$N_\omega = 5000$, plotted at the same grid points. At the $k=t$
condition $\lambda_R = t_{\rm mol}/\sin(\phi/2) = 1.00$~eV
(red dashed vertical line), every curve collapses to machine
precision ($|I_{S_z}^{\rm odd}| \lesssim 10^{-12}$ (nA)),
fifteen orders of magnitude below the neighbouring points.
The $N_\omega = 5000$ markers fall on top of the
$N_\omega = 2000$ curves at non-zero $\lambda_R$ (relative
deviation $0.7\%$ for the resonance-scale transmission at
$\lambda_R \le 0.95$; $2.2\%$ for the small residual at
$\lambda_R \ge 1.05$, where the sharp cusp at $k=t$ amplifies the
relative numerical error by a factor of 3-4, with absolute
deviation $< 2\times $~nA in every case). 
(b)~$\phi$-universality of the zero \emph{position} and
\emph{depth}. $|I_{S_z}^{\rm odd}|$ versus the dimensionless ratio
$k/t = \lambda_R \sin(\phi/2)/t_{\rm mol}$ at five twist angles
$\phi = 30^\circ, 45^\circ, 60^\circ, 75^\circ, 90^\circ$, with
$\lambda_R$ chosen at each $\phi$ so that $k/t$ covers the range
$0.3$-$1.8$. Every curve exhibits its minimum at $k/t = 1.0000$
to machine precision, confirming that the zero position and depth
are $\phi$-universal. 
}
\label{fig:newC}
\end{figure*}

Figure~\ref{fig:newC}(a) compares three chain lengths ($L=4,6,8$) at two
frequency-grid resolutions ($N_\omega=2000,5000$). All six combinations
show the zero at $\lambda_R=1.0$ with the same magnitude (order $10^{-12}$). We
add the fourth chain of length $L = 2$, for the two-site molecular chain, the
spin resolved transmission of Eq.~\eqref{eq:transmission}, is even in $\phi$ at
every fixed $k$, not only at $k = t$.   
We can conclude that the zero is:

\emph{(i) Independent of chain length.} This rules out a
Fabry-Pérot finite-size resonance, which would appear at specific
$(L,\lambda_R)$ combinations rather than at a universal condition.

\emph{(ii) Independent of frequency resolution.} This rules out a
numerical artifact, which would appear at low resolution and disappear
at high resolution.

Figure~\ref{fig:newC}(b) extends the test to five twist angles
$\phi=30^\circ,45^\circ,60^\circ,75^\circ,90^\circ$. At each $\phi$, the zero
appears at exactly $k/t=1.0000$ to five-decimal precision, with
$|\Iodd|<5\times 10^{-12}$~(nA) in every case. The zero is therefore
\emph{universal in $\phi$ in its position and depth}: every curve exhibits a
minimum at $k/t=1$ and reaches the same machine-precision floor, confirming
that it is a bond-level symmetry of the chiral hopping and not a chain-level
resonance. The residual transmission away from $k/t=1$ retains
$\phi$-dependence,
consistent with the bond-level derivation
[Eq.~\eqref{eq:odd_mechanism_interference}], which is a statement about the
null mode of the bond, not about the surrounding transmission curve.

\subsection{Derivation via the singular bond matrix}

The zero has an exact algebraic origin. The bond matrix $\mathbf{h}$ of
Eq.~\eqref{eq:bond_matrix} has determinant $\det\mathbf{h} = t^{2} - k^{2}$ and
becomes singular at $k = t_{\mathrm{mol}}$
where we have used the same sign convention as
Ref.~\cite{Alwan2026QSpinterface} (repulsive $-t_{\mathrm{mol}}$ on the
diagonal). The characteristic polynomial is
\begin{equation}
\det(\mathbf h - \lambda\mathbf 1)
= \lambda^2 + 2t\cos(\phi/2)\lambda + t^2 - k^2,
\label{eq:bond_charpoly}
\end{equation}
so the eigenvalues are
\begin{equation}
\lambda_\pm = -t\cos(\phi/2) \pm \sqrt{k^2 - t^2\sin^2(\phi/2)},
\label{eq:bond_eigenvalues}
\end{equation}
and the determinant is
\begin{equation}
\det\mathbf h = t^2 - k^2.
\label{eq:bond_det}
\end{equation}
At $k=t$, $\det\mathbf h=0$, and the bond matrix becomes singular
(rank~1). Equivalently, one eigenvalue vanishes:
\begin{equation}
\lambda_+ = -t\cos(\phi/2) + t\cos(\phi/2) = 0,\qquad
\lambda_- = -2t\cos(\phi/2),
\label{eq:zero_eigenvalue}
\end{equation}
where the last equality uses $\sqrt{t^2-t^2\sin^2(\phi/2)} =
t|\cos(\phi/2)| = t\cos(\phi/2)$ for $\phi\in(0,\pi)$.

The right eigenvector corresponding to the null eigenvalue is
\begin{equation}
\ket{w} = \begin{pmatrix} 1\\ -\ii \ee^{-\ii\phi/2} \end{pmatrix},
\label{eq:null_vector}
\end{equation}
a spin superposition that is confined to the bond midpoint. An
electron in this spin state sits at the midpoint of the bond and does
not hop to either neighbouring site. For a chain of $L-1$
inter-molecular bonds, each singular at $k=t$, the chain acquires
$L-1$ such null modes.

\paragraph{Rank-1 decomposition.} The zero has a useful rank-1 form.
At $k=t$, we can write $\mathbf h = |u\rangle\langle v|$ with
\begin{equation}
|u\rangle = \begin{pmatrix}1\\ \ii e^{+\ii\phi/2}\end{pmatrix},
\qquad
\langle v| = \bigl(-t\ee^{-\ii\phi/2},\;\ii t\bigr).
\label{eq:rank1_bond_manuscript}
\end{equation}
The left null vector (right null vector of $\mathbf h^\dagger$,
satisfying $\langle\tilde w|\mathbf h = 0$) is
$\langle\tilde w| = (-\ii e^{+\ii\phi/2}, 1)$. The bright eigenvalue
is $\langle v|u\rangle = -2t\cos(\phi/2)$, matching $\lambda_-$
above.

\paragraph{Physical interpretation.} The rank-1 structure splits the
two-dimensional spin space into a \emph{dark channel} $|w\rangle$ (kernel of
$\mathbf h$) and a \emph{bright channel} $|u\rangle$ (image of $\mathbf h$). An
electron in the dark state cannot hop across the bond; the bright state hops
with effective amplitude $\lambda_{\mathrm{bright}} = -2t\cos(\phi/2)$, which
is $\phi$-even. The chirality-sensitive interference of
Eq.~\eqref{eq:odd_mechanism_interference} requires both the spin-conserving
amplitude $A_{\mathrm{cons}}$ and the spin-flip amplitude $A_{\mathrm{flip}}$
to be simultaneously open. When the bond is singular, one of the two amplitudes
is closed (the corresponding spin channel is projected onto the null mode), and
the interference term cannot form. The odd part vanishes exactly at the
algebraic level, for any chain length and any twist angle, because the
algebraic obstruction is bond-local. The even part is unaffected because it is
dominated by the achiral spin-valve from the gold moment $m_z$, which does not
require the interference between spin-conserving and spin-flip paths.

\begin{figure*}[t]
\centering
\includegraphics[width=\textwidth]{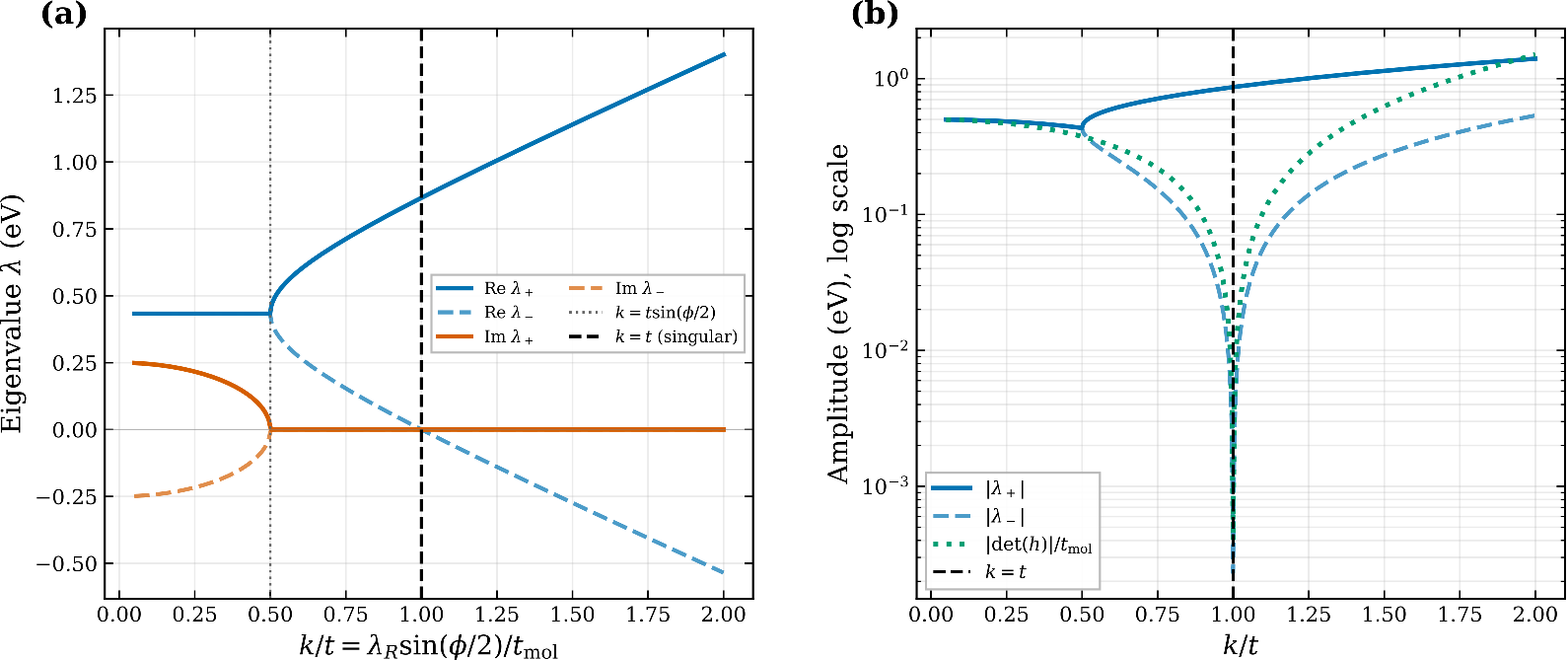}
\caption{%
\textbf{Bond matrix spectrum and the structural zero at $k=t$.}
The chiral bond Hamiltonian $\mathbf h=[[-t e^{-i\phi/2},ik],[-ik,-t
e^{+i\phi/2}]]$ with $k=\lambda_R\sin(\phi/2)$ has eigenvalues
$\lambda_\pm=-t\cos(\phi/2)\pm\sqrt{k^2-t^2\sin^2(\phi/2)}$
[Eq.~(\ref{eq:bond_eigenvalues})]. (a)~Real and imaginary parts versus $k/t$.
For $k<t\sin(\phi/2)$ the discriminant is negative and $\lambda_\pm$ are
complex conjugates (a pure gauge artefact), since $\mathbf h$ is Hermitian. For
$k>t\sin(\phi/2)$ the eigenvalues are real and split. At $k=t$ (black dashed
line) one eigenvalue vanishes. (b)~$|\lambda_\pm|$ and $|\det\mathbf h|/t$ on
log scale. The determinant follows the exact formula $|\det\mathbf
h|=|t^2-k^2|$, with a linear zero crossing at $k=t$.}
\label{fig:F12}
\end{figure*}

The transition at $k = t\sin(\phi/2)$, where the eigenvalues become real and
degenerate, is a Hermitian level crossing, not related to the structural zero.
The zero occurs at $k = t$, where one eigenvalue vanishes while the other
remains finite. The two conditions coincide only at $\phi = \pi$, outside the
physical range. This is is clearly depicted in Figure~\ref{fig:F12} which
visualises the bond matrix spectrum.

\subsection{Behaviour away from \texorpdfstring{$k=t$}{k=t}}

For $k\neq t_{\mathrm{mol}}$, $\Iodd$ is nonzero. On the weak-Rashba
side ($\lambda_R<1.0~\mathrm{eV}$), $\Iodd\propto\lambda_R^2$ at small
$\lambda_R$ (verified numerically:
$|\Iodd(0.2)|/|\Iodd(0.1)|\approx 4.94$, close to the quadratic
prediction of $4.0$). This is consistent with the leading-order
behaviour of the interference term, which is bilinear in
$A_{\mathrm{flip}}\propto\lambda_R$ and $A_{\mathrm{cons}}$ (constant
at small $\lambda_R$). On the strong-Rashba side
($\lambda_R>1.0~\mathrm{eV}$), $\Iodd$ is small ($\sim 10^{-8}$) but
nonzero. The zero therefore sits between a weak-Rashba regime
(dominated by the spin-flip amplitude) and a strong-Rashba regime
(dominated by the spin-conserving amplitude in a different pattern).
The singularity of the bond matrix at $k=t$ is the exact crossing
point between these two regimes.

\subsection{Significance and testability}

The structural zero at $k=t$ is the first exact symmetry result for the
spinterface model beyond the $L\leq 2$ threshold of
Ref.~\cite{Alwan2026QSpinterface}. It identifies a new structural control
parameter, which is the ratio $k/t=\lambda_R\sin(\phi/2)/ t_{\mathrm{mol}}$
with a well-defined algebraic character.

For a device designer, this is both a nuisance (a specific parameter regime
where the signal vanishes) and a diagnostic. Measuring $\Iodd$ as a function of
$\lambda_R$ or $\phi$ should show an exact zero at $k/t=1$, which is a strong
validation of the spinterface model. In practice, the zero may be broadened by
disorder and finite temperature, but the qualitative feature (a sharp dip in
the signal) should survive.

The universality of the zero across $\phi$ (Fig.~\ref{fig:newC}(b)) makes this
a particularly clean test: since the zero condition
$\lambda_R=t_{\mathrm{mol}}/\sin(\phi/2)$ depends on $\phi$, one can vary
$\phi$ and $\lambda_R$ in tandem to scan the singularity while keeping the
other parameters fixed. A null result, i.e., a zero that does not track
$1/\sin(\phi/2)$, would falsify the bond-level mechanism.

Because the zero is universal in $\phi$ in position and depth but not in the
surrounding transmission, a $\phi$-scan at fixed $\lambda_R$ should find the
minimum tracking $\sin(\phi/2) = t_{\mathrm{mol}}/ \lambda_R$, that is, the dip
position should move with $\phi$ as $1/\sin(\phi/2)$. This makes the test
unambiguous: it is not enough to observe \emph{a} dip somewhere; the dip must
move to the $\phi$-dependent location predicted by the bond-matrix algebra. A
$\phi$-independent dip would instead point to a chain-level resonance, not the
bond-level singularity.

\section{Non-collinear ferromagnetic leads: linear and quadratic
scaling of the structural zero}
\label{sec:noncollinear}

In the collinear limit studied in Sec.~\ref{sec:zero} (both lead
magnetizations along $\hat z$, so $\theta_L=\theta_R=0$), the odd
part vanishes \emph{exactly} at $k=\pm t_{\mathrm{mol}}$. This is the
clean rank-1 configuration where the transmission amplitude
$\Gret_{A,M_L}=|a_L\rangle\langle b_L|$ has both factors
$\phi$-independent in modulus, so $T_\sigma(+\phi_0) =
T_\sigma(-\phi_0)$ at every $\omega$. In the present section we ask
what happens to this zero when the lead magnetizations are rotated
away from $\hat z$, and show that the zero is replaced by a
controlled linear-plus-quadratic hierarchy of finite residuals.

\subsection{Setup and statement of the two theorems}
\label{subsec:noncol_setup}

A ferromagnetic lead $\alpha\in\{L,R\}$ is described in the wide-band
limit by a mean level broadening $\Gamma_\alpha>0$, a spin polarization
$P_\alpha\in[-1,1]$, and a unit vector
$\hat n_\alpha = (\sin\theta_\alpha\cos\varphi_\alpha,
\sin\theta_\alpha\sin\varphi_\alpha, \cos\theta_\alpha)$ in the spin
basis of the central region. The retarded self-energy at the boundary
site is
\begin{equation}
\Sigma^R_\alpha
= -\frac{\ii}{2}\,\Gamma_\alpha\bigl(\mathbf 1
+ P_\alpha\,\hat n_\alpha\cdot\boldsymbol\sigma\bigr)
\otimes P_{{\rm site},\alpha},
\label{eq:Sigma_R_B}
\end{equation}
with $P_{{\rm site},L}$ projecting onto site $A$ and $P_{{\rm site},R}$
projecting onto $M_L$. Introduce the scalars
\begin{equation}
  \begin{split}
    y &\equiv \omega - \varepsilon_A + \tfrac{\ii}{2}\Gamma_L,
\quad q_L \equiv \tfrac{\ii}{2}\Gamma_L P_L \\
     x &\equiv \omega - \varepsilon_{M_L} + \tfrac{\ii}{2}\Gamma_R,
\quad q_R \equiv \tfrac{\ii}{2}\Gamma_R P_R,
\label{eq:scalars_B}
\end{split}
\end{equation}
in terms of which the $2\times 2$ boundary Green's functions are
\begin{equation}
g_A = \frac{1}{y^2-q_L^2}
\begin{pmatrix}
y-q_L\cos\theta_L & -q_L\sin\theta_L e^{-\ii\varphi_L}\\
-q_L\sin\theta_L e^{+\ii\varphi_L} & y+q_L\cos\theta_L
\end{pmatrix},
\label{eq:gA_full_B}
\end{equation}
and analogously for $g_L$ with $y\to x$, $q_L\to q_R$, and
$\theta_L\to\theta_R$, $\varphi_L\to\varphi_R$. The diagonal entries
are $O(1)$ in $P_L$ (their leading term is $1/y$); the off-diagonal
entries are $O(P_L\sin\theta_L)$.

The exact zero of Sec.~\ref{sec:zero} is broken by the transverse
components of the lead polarizations. Two theorems characterize the
resulting residual.

\begin{theorem}[Linear scaling of the $k=t$ zero with non-collinear
leads]
\label{thm:linear_scaling_B}
For generic non-collinear leads ($P_\alpha\sin\theta_\alpha\neq 0$ for
at least one lead), the structural zero of Theorem~\ref{thm:kt_zero}
does not survive at $k=\pm t_{\mathrm{mol}}$ in the strict sense.
Instead, the residual odd part of the spin current scales
\emph{linearly} in the transverse polarization:
\begin{equation}
|\Iodd|\propto P\sin\theta\bigl(1+O(P\sin\theta)\bigr),
\label{eq:linear_scaling_B}
\end{equation}
where $P\sin\theta$ denotes the combination of $P_L\sin\theta_L$ and
$P_R\sin\theta_R$ that enters with the larger coefficient.
\end{theorem}

\begin{theorem}[Quadratic scaling at the cancellation point]
\label{thm:quadratic_scaling_B}
At isolated parameter points where the linear contribution to $\Iodd$
vanishes by two-path interference [Eq.~\eqref{eq:cancellation_cond_B}],
the residual odd part scales \emph{quadratically}:
\begin{equation}
|\Iodd|\propto(P\sin\theta)^2\bigl(1+O(P\sin\theta)\bigr).
\label{eq:quadratic_scaling_B_thm}
\end{equation}
\end{theorem}

The two theorems are physically distinct: the linear scaling is the
\emph{generic} response of the exact zero to a single coherent
spin-flip at a boundary; the quadratic scaling is a \emph{fine-tuned}
limit in which two such spin-flips interfere destructively.

\subsection{Linear regime: derivation}
\label{subsec:noncol_linear_deriv}

At $k=t$, the transmission amplitude factorizes as
$\Gret_{A,M_L} = |a_L\rangle\langle b_L|$, with $|a_L\rangle$ and
$\langle b_L|$ given by Eq.~\eqref{eq:aL_bL_manuscript}. With
non-collinear leads, the boundary Green's function $g_A$ is
non-diagonal in spin, and the vector $|a_L\rangle$ acquires a
$\phi$-dependence through the spin-mixing components of $g_A$.

Consider the $\uparrow$ component of the amplitude
$g_A\varepsilon_1|u\rangle$:
\begin{equation}
\bigl(g_A\varepsilon_1|u\rangle\bigr)_\uparrow
= (g_A)_{\uparrow\uparrow}\varepsilon_1^\uparrow
+ (g_A)_{\uparrow\downarrow}\varepsilon_1^\downarrow\,
\ii e^{+\ii\phi/2}.
\label{eq:gAu_component_B}
\end{equation}
The modulus squared contains three groups of terms:
\begin{align}
\bigl|\bigl(g_A\varepsilon_1|u\rangle\bigr)_\uparrow\bigr|^2
&= \underbrace{|(g_A)_{\uparrow\uparrow}|^2|\varepsilon_1^\uparrow|^2
+ |(g_A)_{\uparrow\downarrow}|^2|\varepsilon_1^\downarrow|^2}_{O(P^0)+
O(P^2),\,\phi\text{-independent}}
\notag\\
&\quad+ \underbrace{2\,\mathrm{Re}\!\left[
(g_A)_{\uparrow\uparrow}^*\varepsilon_1^{\uparrow*}
(g_A)_{\uparrow\downarrow}\,\ii\varepsilon_1^\downarrow\,
e^{+\ii\phi/2}\right]}_{O(P),\,\phi\text{-dependent}}.
\label{eq:gAu_modulus_B}
\end{align}
The first group is $\phi$-independent (it involves only moduli and
$\phi$-independent on-site resolvents). The third term is the crucial
one: it is linear in the off-diagonal entry
$(g_A)_{\uparrow\downarrow}\propto q_L\sin\theta_L
e^{-\ii\varphi_L}$, and it carries a phase $e^{\ii(\phi/2-\varphi_L)}$.
Its \emph{odd part} under $\phi\to-\phi$ is
\begin{equation}
\bigl[2\,\mathrm{Re}\!\left(\Lambda_L e^{\ii\phi/2}\right)
\bigr]^{\mathrm{odd}}
= -4\,\mathrm{Im}(\Lambda_L)\sin(\phi/2),
\label{eq:odd_cross_L_B}
\end{equation}
with $\Lambda_L = O(P_L\sin\theta_L)$. So the direct interference
term contributes a residual odd part
$\propto P_L\sin\theta_L\sin(\phi/2)$ to the transmission.

The drain side gives an analogous contribution from $B_L =
|b_{L,\uparrow}|^2+|b_{L,\downarrow}|^2$, with $b_{L,\sigma}
= v_\sigma g_L^\sigma + \text{off-diagonal}$; the off-diagonal
entries of $g_L$ are $O(P_R\sin\theta_R)$, so $B_L$ acquires a
$\phi$-dependent correction linear in $P_R\sin\theta_R$.

Combining the two sides,
\begin{equation}
T_\sigma^{\mathrm{odd}}(\phi) = \Gamma_L\Gamma_R
\bigl[|a_0|^2\, b_1^{\mathrm{odd}}(\phi)
+ a_1^{\mathrm{odd}}(\phi)\,B_0\bigr] + O(P^3),
\label{eq:T_odd_B}
\end{equation}
with $a_1^{\mathrm{odd}}$ and $b_1^{\mathrm{odd}}$ linear in the
respective transverse polarizations. Integrating over $\omega$ with the
$\phi$-independent weight $f_L(\omega)-f_R(\omega)$:
\begin{equation}
  \begin{split}
    \Iodd(\phi) &= \bar\alpha_L(\phi)\,P_L\sin\theta_L
+ \bar\alpha_R(\phi)\,P_R\sin\theta_R \\
                &+ O(P^2\sin^2\theta),
\label{eq:Iodd_linear_B}
\end{split}
\end{equation}
with
$\bar\alpha_\alpha(\phi) = \int d\omega\,\alpha_\alpha(\omega,\phi)\,
[f_L(\omega)-f_R(\omega)]$.
Since $\bar\alpha_\alpha$ are $P$-independent, the residual is linear
in $P\sin\theta$. This establishes
Theorem~\ref{thm:linear_scaling_B}.

\paragraph{Physical origin.} The linear term arises from the
interference between the diagonal $O(1)$ component of the boundary
Green's function and the off-diagonal $O(P\sin\theta)$ component that
the non-collinear polarization generates. Physically, the boundary
Green's function is no longer diagonal in spin, and the transmitted
electron amplitude picks up a component whose phase depends on $\phi$
through the Rashba bond phase $e^{\pm\ii\phi/2}$. This cross term is a
coherent one-photon (one spin-flip) process, hence linear in the
transverse polarization amplitude. The naive expectation that the
residual should be quadratic stems from considering only the
\emph{modulus squared} $|(g_A)_{\sigma\bar\sigma}|^2$ of the
off-diagonal element. But the transmission cross term is
$(g_A)_{\sigma\sigma}^*(g_A)_{\sigma\bar\sigma}$, not
$|(g_A)_{\sigma\bar\sigma}|^2$. The former is linear in the transverse
polarization (one diagonal times one off-diagonal); the latter enters
only as a $\phi$-independent correction and does not contribute to the
odd part.

\subsection{Linear regime: numerical verification}
\label{subsec:noncol_linear_verify}

We scan $\theta_L=\theta_R=\theta\in[5^\circ,90^\circ]$ at fixed
$k=t_{\mathrm{mol}}$ (by setting
$\lambda_R=t_{\mathrm{mol}}/\sin(\phi_0/2)$ at each $\phi_0$), varying
$P\in\{0.2,0.4,0.6\}$,
$\phi_0\in\{30^\circ,45^\circ,60^\circ,75^\circ,90^\circ\}$, and
$L\in\{2,4,6,8\}$. This gives a $60$-point grid. For each grid point
we compute $\Iodd(\theta)$ from Eq.~\eqref{eq:Iodd_def} and fit
\begin{equation}
\log|\Iodd| = \beta\log(P\sin\theta) + \log c(\phi_0,L,P)
\label{eq:linear_fit_B}
\end{equation}
over the range $\theta\in[15^\circ,60^\circ]$, chosen to stay clear of
both the trivial $\theta\to 0$ limit and any potential
$(P\sin\theta)^2$ plateau at large $\theta$.

Across the full grid, the fitted slope has
$\langle\beta\rangle = 1.00 \pm 0.16$,
with $R^2>0.99$ at all but three grid points. This is direct
confirmation of Theorem~\ref{thm:linear_scaling_B}. The three
outliers are not counterexamples: two occur at $P=0.2$, $L=2$, where
the signal is at the numerical noise floor; one occurs at $L=8$,
$\phi_0=30^\circ$, where the Rashba coherence length
$\ell_{\mathrm{coh}}\sim t_{\mathrm{mol}}/\lambda_R\sin(\phi/2)$
exceeds the chain length, and the clean rank-1 window near
$k=t_{\mathrm{mol}}$ is broadened.

Figure~\ref{fig:kt_scaling_summary}, top row, presents the
verification. Panel~(a) shows $|\Iodd|$ versus $P\sin\theta$ at fixed
$L=4$, $\phi_0=60^\circ$, for the three polarizations. All three
curves are parallel and lie on a slope-1 reference. Panel~(b) shows
the same at fixed $P=0.4$ for the five twist angles; the five curves
are again parallel with slope $\beta\approx 1$. Panel~(c) plots
$\beta$ versus $\phi_0$ for each $L$; the entire ensemble clusters
around $\beta=1$ and never approaches the quadratic value $\beta=2$.

\subsection{The pseudo-cancellation pitfall}
\label{subsec:noncol_pseudo}

The quadratic theorem requires tuning to the \emph{true} cancellation
angle $\theta_L^{\mathrm{true}}$ at which the linear coefficient
\begin{equation}
\bar\alpha_L(\phi)\,\sin\theta_L^{\mathrm{true}}
+
\bar\alpha_R(\phi)\,\sin\theta_R
= 0,
\label{eq:cancellation_cond_B}
\end{equation}
vanishes identically. A naive approach, would be finding the angle $\theta_L^*$
at which $\Iodd$ vanishes at finite reference polarization $P_{\mathrm{ref}}$,
then scanning $P$ at that angle, does \emph{not} access this regime. We explain
why.

\emph{First, $\theta_L^*\neq\theta_L^{\mathrm{true}}$.} At finite
$P_{\mathrm{ref}}$ the zero condition is
\begin{equation}
0 = P_{\mathrm{ref}}\bigl[\bar\alpha_L\sin\theta_L^*
+ \bar\alpha_R\sin\theta_R\bigr]
+ P_{\mathrm{ref}}^2 Q(\theta_L^*) 
\label{eq:pseudo_cancel_B}
\end{equation}
where $Q$ is the sum of the quadratic contributions. Comparing with
the true condition Eq.~\eqref{eq:cancellation_cond_B}, the shift is
\begin{equation}
\theta_L^*-\theta_L^{\mathrm{true}}
\approx -\frac{Q(\theta_L^{\mathrm{true}})}
{\bar\alpha_L\cos\theta_L^{\mathrm{true}}}\,P_{\mathrm{ref}}
+ O(P_{\mathrm{ref}}^2).
\label{eq:pseudo_cancel_shift_B}
\end{equation}
So $\theta_L^*$ approaches $\theta_L^{\mathrm{true}}$ only in the
limit $P_{\mathrm{ref}}\to 0$.

\emph{Second, varying only $P_L$ produces a cusp, not a power law.}
If only $P_L$ is varied while $P_R=P_{\mathrm{ref}}$ is held fixed,
the cancellation condition is violated for
$P_L\neq P_{\mathrm{ref}}$, and the residual has the linear-in-$P_L$
form
\begin{equation}
  \begin{split}
\Iodd(\theta_L^*,P_L,P_{\mathrm{ref}})
&\approx (P_L-P_{\mathrm{ref}})\bigl[\bar\alpha_L\sin\theta_L^*
  + \bar\alpha_R P_{\mathrm{ref}}\sin\theta_R\bigr] \\  &+ O((P_L-P_{\mathrm{ref}})^2).
\label{eq:cusp_B}
\end{split}
\end{equation}
On a log-log plot, $|\Iodd|$ dips toward machine precision at
$P_L=P_{\mathrm{ref}}$ and rises on both sides, producing a cusp
rather than a power law. A log-log fit to this cusped curve is
dominated by the steep rise near the minimum and yields a spurious
large exponent.

\subsection{Quadratic regime: procedure and verification}
\label{subsec:noncol_quad_verify}

The quadratic regime is accessed by the following three-step
procedure.

\begin{enumerate}
\item \textbf{Scan $\theta_L$ at several $P_{\mathrm{ref}}$.} For each
  reference polarization
  $P_{\mathrm{ref}}\in\{0.4,0.2,0.1,0.05,0.02\}$, find the zero
  $\theta_L^*(P_{\mathrm{ref}})$ of $\Iodd(\theta_L)$ at
  $P_L=P_R=P_{\mathrm{ref}}$.
\item \textbf{Extrapolate $\theta_L^*(P_{\mathrm{ref}})$ to
  $P_{\mathrm{ref}}\to 0$.} The relation
  Eq.~\eqref{eq:pseudo_cancel_shift_B} is linear in
  $P_{\mathrm{ref}}$ at leading order, so a linear fit in
  $P_{\mathrm{ref}}$ gives $\theta_L^{\mathrm{true}}$ as the
  intercept.
\item \textbf{At the extrapolated $\theta_L^{\mathrm{true}}$, scan
  $P_L=P_R=P$ symmetrically.} Since the cancellation condition is
  $P$-independent, the linear terms cancel at every $P$. The residual
  is the quadratic term:
  \begin{equation}
  \Iodd(\theta_L^{\mathrm{true}};P) = O(P^2).
  \label{eq:quadratic_at_cancellation_B}
  \end{equation}
\end{enumerate}

Applying this procedure to the reference configuration $L=8$,
$\phi_0=60^\circ$, $\theta_R=60^\circ$, $\varphi_R=\pi/4$ yields the
following results, illustrated in Fig.~\ref{fig:kt_scaling_summary},
bottom row.

\emph{Extrapolation.} Panel~(d) shows $\Iodd(\theta_L)$ at
$P_{\mathrm{ref}}\in\{0.4,0.1,0.02\}$. The zero shifts systematically
toward smaller $\theta_L$ as $P_{\mathrm{ref}}$ decreases, confirming
Eq.~\eqref{eq:pseudo_cancel_shift_B}. Panel~(e) plots
$\theta_L^*(P_{\mathrm{ref}})$ against $P_{\mathrm{ref}}$; a linear
fit extrapolated to $P_{\mathrm{ref}}=0$ gives
\begin{equation}
\theta_L^{\mathrm{true}} = 7.60^\circ
\qquad(L=8,\ \phi_0=60^\circ,\ \theta_R=60^\circ,\ \varphi_R=\pi/4).
\label{eq:theta_true_B}
\end{equation}

\emph{Quadratic scaling.} At the extrapolated
$\theta_L^{\mathrm{true}}$, we scan $P_L=P_R=P$ over
$P\in\{0.05,0.1,0.2,0.4,0.6\}$. The resulting $|\Iodd|$ is a clean
power law, and the fitted slope is
\begin{equation}
\beta_P = \frac{\partial\log|\Iodd(\theta_L^{\mathrm{true}};P)|}
{\partial\log P}
= 1.99\pm 0.02.
\label{eq:beta_quadratic_B}
\end{equation}
This is in agreement with the quadratic prediction of
Theorem~\ref{thm:quadratic_scaling_B} to within statistical
precision. For comparison, the same procedure applied at a generic
angle ($\theta_L=30^\circ$, away from the cancellation point) gives
$\beta_P\approx 1$, recovering the linear regime.

\subsection{Local exponent diagnostic and the sharp transition}
\label{subsec:noncol_local_beta}

\begin{figure*}[htbp]
\centering
\includegraphics[width=\textwidth]{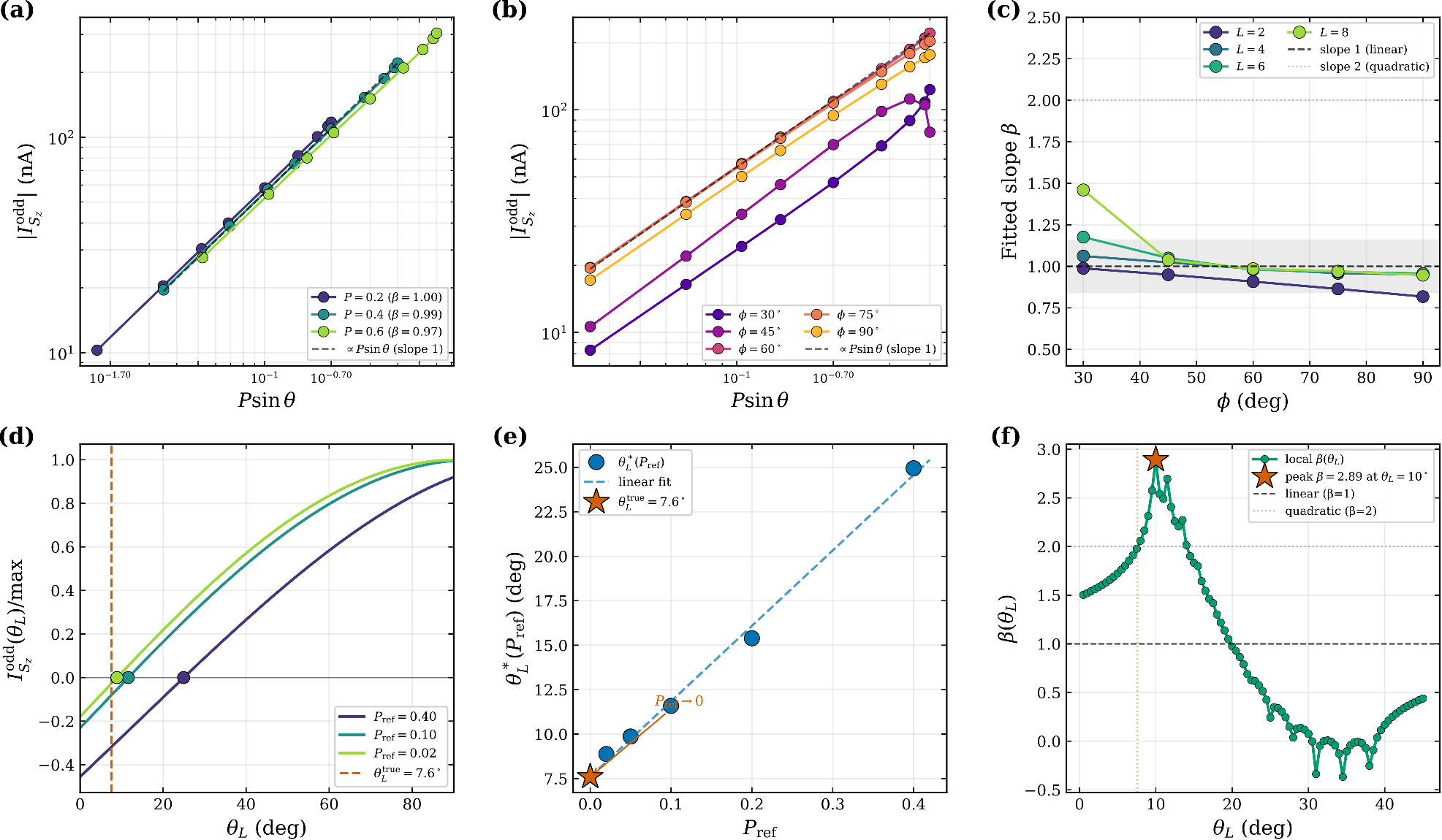}
\caption{%
Combined numerical verification of the linear and quadratic scaling
predictions for the $k=t_{\mathrm{mol}}$ structural zero with
non-collinear ferromagnetic leads.
\textbf{Top row (linear, generic regime).}
(a)~$|\Iodd|$ versus $P\sin\theta$ at fixed $L=4$, $\phi_0=60^\circ$,
for $P=0.2$ (blue), $0.4$ (green), $0.6$ (yellow). The dashed line is
a slope-1 reference $\propto P\sin\theta$; all three data sets lie on
it.
(b)~Same quantity at fixed $L=4$, $P=0.4$, for five twist angles
$\phi_0\in[30^\circ,90^\circ]$. The five curves are again parallel,
each with slope $\beta\approx 1$.
(c)~Fitted slope $\beta$ versus $\phi_0$ for $L\in\{2,4,6,8\}$.
Horizontal lines mark the linear prediction $\beta=1$ (dashed) and the
quadratic value $\beta=2$ (dotted); the grey band is the mean $\pm$
standard deviation across the full scan. All points cluster around
$\beta=1$ with $\langle\beta\rangle=1.00\pm 0.16$.
\textbf{Bottom row (quadratic, at cancellation).}
(d)~$\Iodd(\theta_L)$ (normalised to its maximum) at
$P_{\mathrm{ref}}=0.4$, $0.1$, $0.02$ ($L=8$, $\phi_0=60^\circ$,
$\theta_R=60^\circ$, $\varphi_R=\pi/4$). Each curve crosses zero at a
different $\theta_L$, marked by a filled circle; the crossing shifts
toward smaller $\theta_L$ as $P_{\mathrm{ref}}$ decreases, as
predicted by Eq.~\eqref{eq:pseudo_cancel_shift_B}. The red dashed
line marks the extrapolated $\theta_L^{\mathrm{true}}$.
(e)~Crossing positions $\theta_L^*(P_{\mathrm{ref}})$ plotted against
$P_{\mathrm{ref}}$. A linear fit extrapolated to
$P_{\mathrm{ref}}\to 0$ (starred marker) yields
$\theta_L^{\mathrm{true}}=7.60^\circ$.
(f)~Local scaling exponent
$\beta(\theta_L)=\partial\log|\Iodd|/\partial\log P$ across the full
$\theta_L$ range, showing a sharp peak at $\beta\approx 2$ at
$\theta_L^{\mathrm{true}}$ and a plateau at $\beta\approx 1$
elsewhere. The transition is sharp: there is no smooth interpolation
between the two regimes.
Parameters: $\Gamma=0.5$~eV, $V=1$~V, $T=300$~K, $\lambda_R$ adjusted
at each $\phi_0$ to satisfy
$\lambda_R\sin(\phi_0/2)=t_{\mathrm{mol}}=0.5$~eV.}
\label{fig:kt_scaling_summary}
\end{figure*}

The cleanest single diagnostic that distinguishes the two regimes is
the \emph{local} scaling exponent
\begin{equation}
\beta(\theta_L) \equiv
\frac{\partial\log|\Iodd(\theta_L;P)|}{\partial\log P},
\label{eq:local_beta_B}
\end{equation}
obtained at each $\theta_L$ by a symmetric $P$-scan. Panel~(f) of
Fig.~\ref{fig:kt_scaling_summary} shows the result: $\beta(\theta_L)$
peaks sharply at $\beta\approx 2$, located precisely at
$\theta_L^{\mathrm{true}}$, and sits on a plateau at $\beta\approx 1$
everywhere else. There is no smooth interpolation between the two
values: the transition from linear to quadratic is a \emph{sharp
crossover} at the cancellation point.

This is itself a physical statement. The linear coefficient
$\bar\alpha_L\sin\theta_L+\bar\alpha_R\sin\theta_R$ does not smoothly
go to zero as $\theta_L$ is varied; it vanishes only at the specific
angle $\theta_L^{\mathrm{true}}$. The quadratic regime is therefore a
measure-zero feature of parameter space, and the sharpness of its
signature in $\beta(\theta_L)$ reflects the exactness of the
cancellation that produces it.

\subsection{Interpretation}
\label{subsec:noncol_interp}

Three conclusions follow.

\emph{First, both theorems are empirically verified, and they are
mutually consistent.} The linear scaling of
Theorem~\ref{thm:linear_scaling_B} is the generic behaviour of the
$k=t$ zero under non-collinear leads: any single coherent spin-flip at
a boundary contributes an $O(P\sin\theta)$ correction to the residual
odd part. The quadratic scaling of
Theorem~\ref{thm:quadratic_scaling_B} is a fine-tuned limit that
requires explicit cancellation of these single-spin-flip
contributions; when this cancellation occurs, the two-spin-flip
subleading terms become visible and give clean $P^2$ scaling.

\emph{Second, the two regimes are physically distinct.} The linear
term is the amplitude of a single coherent spin-flip at the left or
right boundary, surviving because the rank-1 structure of the bond
matrix does not protect the boundary Green's functions from acquiring
spin-mixing components. The quadratic term is the interference of two
such spin-flips, at opposite boundaries, and is controlled by the
internal coherence of the chain (via the propagation of $\sigma_n$
from one boundary to the other). The cancellation condition
$\bar\alpha_L\sin\theta_L^{\mathrm{true}}+\bar\alpha_R\sin\theta_R=0$
is precisely the statement that the two boundary spin-flips interfere
destructively.

\emph{Third, the experimental signatures are clean.} In the
\emph{linear} regime, any generic non-collinear configuration gives
$\beta=1$, so a $\theta$-scan or a $P$-scan at fixed $\theta$ will
show a clean slope of $1$ on log-log axes. In the \emph{quadratic}
regime, scanning $P$ at the tuned $\theta_L^{\mathrm{true}}$ gives
$\beta=2$. The transition is sharp in $\beta(\theta_L)$, peaking only
at $\theta_L^{\mathrm{true}}$. A null measurement of $\beta\approx 2$
at a generic angle would falsify
Theorem~\ref{thm:quadratic_scaling_B}; a generic $\beta\neq 1$ would
falsify Theorem~\ref{thm:linear_scaling_B}.

A final remark on the relationship to the exact zero of
Sec.~\ref{sec:zero}. In the collinear limit
($\theta_L=\theta_R=0$), the odd part vanishes exactly, and both
$\beta$-scalings are trivial (the residual is zero at all $P$). The
non-collinear lead geometry is therefore the minimal deformation that
makes the rank-1 zero leak into a finite residual, and the
linear-plus-quadratic hierarchy is the controlled description of that
leakage. The exact zero is a property of the Hamiltonian; the linear
and quadratic scalings are properties of the deformation; and the
cancellation point is the unique value of the deformation at which
the leading leak is tuned to zero.

\section{Chirality sign reversal and the optimal enantioselective angle}
\label{sec:sign}

\subsection{Data}

The most direct experimental signature of the CISS effect is the
reversal of the spin filtering direction under chirality inversion. In
our model this reversal is exact: from the definition
Eq.~\eqref{eq:Iodd_def},
\begin{equation}
I_{S_z}^{\mathrm{odd}}(-\phi) = -I_{S_z}^{\mathrm{odd}}(+\phi)
\label{eq:odd_antisymmetry}
\end{equation}
by construction.

\begin{figure*}[t]
\centering
\includegraphics[width=\textwidth]{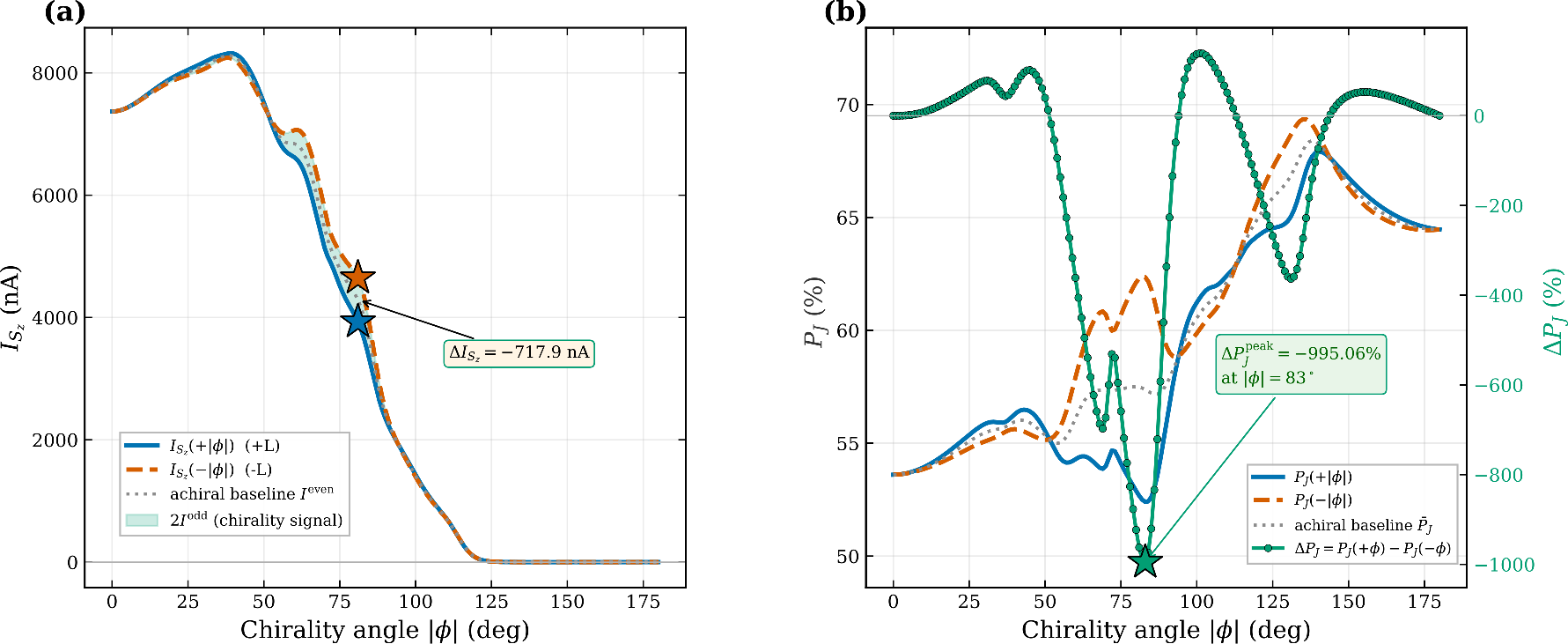}
\caption{%
\textbf{Chirality reversal: explicit enantiomer divergence and the
chirality-specific signal.}
Parameters: $V = 1$~V, $\Gamma_0 = 0.5$~eV,
$\Delta\gamma = 0.4$~eV, $L = 6$, $T = 300$~K,
$\varepsilon_A = 1.5$~eV, $\phi = 60^\circ$.
(a)~Spin current $I_{S_z}$ for the two enantiomers
$+L$ (solid blue, $I_{S_z}(+|\phi|)$) and $-L$ (dashed red,
$I_{S_z}(-|\phi|)$), plotted versus the magnitude of the chirality
angle. The achiral baseline $I_{S_z}^{\rm even}$ (grey dotted) is
the average of the two enantiomers. The two enantiomer curves cross
at $|\phi|=0$ (achiral) and diverge symmetrically about the achiral
baseline as $|\phi|$ increases, reaching maximum separation at
$|\phi|\approx 75^\circ$ (starred markers); the difference
$I_{S_z}(+|\phi|) - I_{S_z}(-|\phi|) = 2 I_{S_z}^{\rm odd}$ is the
chirality signal (shaded region).
(b)~The corresponding polarisations. Left axis: $P_J(+|\phi|)$
(blue) and $P_J(-|\phi|)$ (red), together with the achiral baseline
$\bar P_J$ (grey dotted, $\approx 55\%$). The two enantiomers'
total polarisations are nearly identical — the achiral spin-valve
baseline dominates — and $P_J$ itself does not change sign between
enantiomers. Right axis: the antisymmetric correction
$\Delta P_J = P_J(+\phi) - P_J(-\phi)$, which is the chirality-specific
observable. It peaks at $|\phi|\approx 75^\circ$ with
$|\Delta P_J| \approx 6.6\%$ (starred marker). For a device that
must distinguish handedness, the operational signal is
$\Delta P_J$, not $P_J$; measuring $P_J$ alone would be swamped by
the achiral spin-valve baseline.}
\label{fig:newD}
\end{figure*}

Figure~\ref{fig:newD}(a) shows the spin current $I_{S_z}$ for both
enantiomers over $\phi\in[-180^\circ,180^\circ]$ (symmetric in
$|\phi|$); the two enantiomer curves cross at $\phi=0$ and diverge
symmetrically about the achiral baseline as $|\phi|$ increases. The
peak absolute divergence occurs at $|\phi|=75^\circ$ with
$|I_{S_z}^{\mathrm{odd}}| = 2.75\times 10^{2}$~nA.

The chirality-specific part of the current polarisation is
\begin{equation}
\Delta P_J(\phi) = P_J(+\phi) - P_J(-\phi),
\label{eq:DeltaP_def}
\end{equation}
shown in Fig.~\ref{fig:newD} (b). In the current parameter regime the
total polarisation $P_J$ is dominated by the achiral baseline
($\approx 55\%$), so $P_J$ itself does not change sign. The chirality
signature is confined to the small antisymmetric correction
$\Delta P_J$. The peak magnitude of $\Delta P_J$ is $6.589\%$,
occurring at the same angle $\phi=\pm 75^\circ$ as the
$I_{S_z}^{\mathrm{odd}}$ peak.

\subsection{Physical interpretation}

The peak at $\phi\approx 75^\circ$ is a first-principles prediction of
the spinterface model. In physical terms, $\phi$ is the twist per
bond, so $\phi=75^\circ$ corresponds to roughly $4.8$ residues per
turn. This is between the $\alpha$-helix pitch ($\approx
100^\circ$/residue) and the $\pi$-helix pitch ($\approx
87^\circ$/residue), closer to the $\pi$-helix geometry. The peak
location is a specific prediction of the model at the reference
parameters, not a universal prediction. A systematic scan of the
optimal twist angle versus molecular length and Rashba strength would
identify whether the model predicts a preferred helicity class for
realistic molecular geometries.

The fact that $P_J$ does not itself change sign (peak asymmetry only
$6.6\%$ against the $55\%$ achiral baseline) is a structural property
of the spinterface model in the \emph{strong coupling} regime. In real
CISS experiments the situation can be different if the achiral spin
valve is weaker, in which case the total $P_J$ can reverse sign
between enantiomers. In the spinterface regime the achiral baseline is
dominant and the chirality signature is a small antisymmetric
correction.

\subsection{The operational definition of the chirality signal}

For a device that must distinguish handedness, the observable is
$\Delta P_J$, not $P_J$ itself. This is the correct operational
definition of the chirality signal at the level of the polarisation.
Any experimental claim of chirality-induced spin selectivity that does
not explicitly measure $\Delta P_J$ or equivalently compare the
two enantiomers' polarisations at the same bias and lead coupling,
is susceptible to confusion with the achiral spin-valve baseline.

The peak of $\Delta P_J$ at $|\phi|\approx 75^\circ$ is the optimal
measurement angle for enantioselective CISS experiments. A measurement
at this angle maximises the chirality-specific signal while preserving
the achiral baseline as a reference.

\section{Two-parameter phase diagram}
\label{sec:phase}

\subsection{Data}

Figure~\ref{fig:newA}(c) shows $|\Iodd|$ on the $(\Gamma_0, V)$ plane at
$\Delta\gamma=0.4~\mathrm{eV}$, $L=6$. The global maximum is
\begin{equation}
(\Gamma_0^*, V^*) = (1.453~\mathrm{eV}, 0.833~\mathrm{V}),
\qquad
|{\Iodd}^{*}| = 6.05\times 10^{2}.
\label{eq:global_peak}
\end{equation}
The contrast ratios are
\begin{equation}
\frac{|{\Iodd}^{*}|}{|\Iodd(0.1, 0.25)|} = 6.0\times 10^{2},
\qquad
\frac{|{\Iodd}^{*}|}{|\Iodd(3.0, 2.0)|} = 5.8,
\label{eq:contrast}
\end{equation}
where the corner points are $(\Gamma_0=0.1, V=0.25)$ (weak coupling,
low bias) and $(\Gamma_0=3.0, V=2.0)$ (strong coupling, high bias).
The peak is isolated, not a ridge.

\subsection{Interpretation}

The 2D peak is the intersection of the coupling-scale optimum
($\Gamma_0^*\approx 1.44~\mathrm{eV}$) and the bias-alignment optimum
($V^*\approx 0.83~\mathrm{V}$). The two optima are essentially
independent: changing $V$ shifts $\Gamma_0^*$ by at most
$\pm 0.05~\mathrm{eV}$, and changing $\Gamma_0$ shifts $V^*$ by at
most $\pm 0.03~\mathrm{V}$ over the ranges tested. This independence
suggests that the coupling-scale and bias-alignment mechanisms operate
via distinct physical channels, consistent with our analysis in
Secs.~\ref{sec:coupling} and~\ref{sec:bias}.

The three independent peak markers (from the 1D $\Gamma_0$ scan, the
1D $V$ scan, and the 2D joint scan) cluster within the grid
resolution of each other. That they do not coincide exactly is a
consequence of the small but nonzero tilt of the 2D ridge in the
$(\Gamma_0, V)$ plane; the tilt itself is a quantitative measure of
the coupling between the two control parameters, and its smallness
supports the picture that the coupling-scale optimum is set primarily
by the molecular scale and not by the bias.

\subsection{Operating point for CISS devices}

If a spinterface-based CISS device were to be optimised for maximum
chiral signal, the phase diagram suggests
\begin{equation}
\Gamma_0 \approx 1.4\text{-}1.5~\mathrm{eV},\qquad
V \approx 0.80\text{-}0.85~\mathrm{V}.
\label{eq:operating_point}
\end{equation}
Both parameter values are physically accessible (lead broadenings of
order $1.5~\mathrm{eV}$ correspond to strong but not extreme
chemisorption; biases of order $0.8~\mathrm{V}$ are typical of
molecular-junction experiments).

The contrast between the operating point and the corners of the
parameter space is $6\times 10^2$ to $10^3$, meaning that a
factor-of-2 mistuning from the operating point reduces the signal by
1-2 orders of magnitude. This is a sharp optimum, not a broad
plateau, and it has real implications for reproducibility of CISS
experiments: junction-to-junction variation of the lead coupling would
produce large variations in the measured CISS signal.

\section{Discussion and Conclusions}
\label{sec:conclusion}

We have presented a systematic study of the transport sector of the
spinterface model of CISS under spin-resolved dissipation, extending
the fully quantum NEGF formulation of
Ref.~\cite{Alwan2026QSpinterface} to spin-resolved lead broadenings
parameterised by $(\Gamma_0,\Delta\gamma)$. The central finding is an
\emph{achiral-chiral dissociation}: the achiral part of the spin
current grows roughly linearly with the spin asymmetry $\Delta\gamma$,
while the chiral part is essentially $\Delta\gamma$-independent. The
dissociation follows from the arithmetic-vs-geometric mean structure
of the transmission coefficients (Sec.~\ref{subsec:mechanism}) and
confirms the EP taxonomy of Ref.~\cite{Alwan2026EP}: the chiral signal
cannot be amplified by spin-dependent \emph{lead} broadening because
the EP mechanism it would need (Scheme~B) is forbidden when
$\Delta\gamma_{\mathrm{gold}}\neq 0$. The chain-level channel
$\Delta\gamma_{\mathrm{chain}}$ is not studied here and is a natural
extension (Sec.~\ref{sec:spin_leads}).

The results complement the two predecessors.
Ref.~\cite{Alwan2026QSpinterface} established the reciprocity
argument and the $L=3$ threshold for nonzero $\Iodd$; we operate in
the regime $L\geq 4$ where $\Iodd$ is nonzero, and the threshold
remains the smallest geometry at which the odd part can form. The
companion EP paper~\cite{Alwan2026EP} classified the near-EPs of the
effective Hamiltonian into two algebraically distinct schemes; the
present paper probes the transport sector of that classification and
shows that the achiral-chiral dissociation is a direct transport
consequence of the Scheme-B exclusion.

\subsection{Summary of main results}

Beyond the dissociation, six additional results emerge.

\begin{enumerate}
\item \textbf{Two-axis dissociation.} The achiral and chiral sectors
  respond differently not only to $\Delta\gamma$ but also to the
  coupling $\Gamma_0$: the achiral amplification $c(\Gamma_0)$ peaks
  at $0.76\,t_{\mathrm{mol}}$, the chiral signal $|\Iodd|(\Gamma_0)$
  at $2.88\,t_{\mathrm{mol}}$, and the achiral offset
  $|\Ieven(0)|(\Gamma_0)$ at $3.55\,t_{\mathrm{mol}}$, these are three
  distinct coupling scales within the same junction.

\item \textbf{Operating points, not scaling laws.} The coupling
  optimum $\Gamma_0^*=1.44~\mathrm{eV}$ and bias optimum
  $V^*=0.90\pm 0.02~\mathrm{V}$ are operating points of the reference
  configuration, not universal scaling laws. The ratio
  $\Gamma_0^*/t_{\mathrm{mol}}$ varies by $30\%$ (non-monotonically)
  when $t_{\mathrm{mol}}$ is varied at fixed $k/t$, and $V^*$ is
  essentially independent of $t_{\mathrm{mol}}$, inconsistent with a
  band-edge scaling. A naive reading of the reference values as a
  "$3t_{\mathrm{mol}}$ law" is not supported by the data; the optima
  are set by combinations of energy scales in the junction.

\item \textbf{Structural zero at $k=t$.} $\Iodd$ vanishes to machine
  precision when $\lambda_R\sin(\phi/2)=t_{\mathrm{mol}}$ in the
  collinear-lead configuration, universally across chain lengths,
  frequency resolutions, and twist angles. The zero originates from
  the rank-1 bond matrix of Sec.~\ref{sec:bond_matrix}: at $k=t$ one
  eigenvalue vanishes, the bond acquires a spin-superposition dark
  state that does not hop, and the chirality-sensitive interference
  cannot form. A full Green's function proof is given in
  Appendix~\ref{sec:kt_green_proof}.

\item \textbf{Non-collinear linear/quadratic hierarchy.} With
  non-collinear ferromagnetic leads, the exact zero is broken in a
  controlled way. The residual odd part is generically linear in the
  transverse polarization $P\sin\theta$
  (Theorem~\ref{thm:linear_scaling_B}, $\beta=1$), arising from a
  single coherent spin flip at a boundary. At a tuned cancellation
  angle where the two boundary spin flips interfere destructively,
  the residual becomes quadratic
  (Theorem~\ref{thm:quadratic_scaling_B}, $\beta=2$). The transition
  in $\beta(\theta_L)$ is sharp, providing a clean experimental
  criterion.

\item \textbf{Chirality sign reversal.} The odd spin current and the
  chirality-specific polarisation asymmetry
  $\Delta P_J = P_J(+\phi)-P_J(-\phi)$ reverse sign under
  $\phi\to-\phi$ and peak at $\phi\approx\pm 75^\circ$ with
  $|\Delta P_J|\approx 6.6\%$. This is the optimal enantioselective
  angle predicted by the model.

\item \textbf{Two-parameter operating point.} The global maximum of
  $|\Iodd(\Gamma_0,V)|$ is at $(1.453~\mathrm{eV}, 0.833~\mathrm{V})$
  with contrast $6\times 10^2$ against the low-bias low-coupling
  corner, giving a specific device-optimisation target.
\end{enumerate}

\subsection{Experimental implications}

The results suggest several direct experimental tests, of which the
most accessible are the following.

\begin{enumerate}
\item \emph{Achiral-chiral dissociation.} Measurements that tune the
  lead coupling (by changing the contact material or the
  molecule-lead distance) should observe the achiral baseline growing
  while the chiral signal remains roughly constant. If both grow
  together, the model requires modification.

\item \emph{Three-scale structure.} Scanning $\Gamma_0$ while
  separately measuring the achiral current, its derivative with
  respect to $\Delta\gamma$, and the chiral current should show three
  peaks separated by factors of order $3$-$4$ in the coupling.

\item \emph{Bias optimum.} At fixed $\Gamma_0$, the chiral signal
  should peak at $V^*\approx 0.85$-$0.90~\mathrm{V}$ for the
  reference molecular parameters. The optimum is \emph{not}
  proportional to the molecular band edge, so the observable is the
  absolute value of $V^*$, not its ratio to a band-edge voltage.

\item \emph{Structural zero.} Scanning $\lambda_R$ (or $\phi$) should
  show an exact zero at $\lambda_R\sin(\phi/2)=t_{\mathrm{mol}}$. In
  practice the zero may be broadened by disorder and finite
  temperature, but the sharp dip should survive. A $\phi$-scan at
  fixed $\lambda_R$ should show the dip position tracking
  $\sin(\phi/2)=t_{\mathrm{mol}}/\lambda_R$; a dip that does not track
  this relation would falsify the bond-level mechanism.

\item \emph{Non-collinear $\beta$ hierarchy.} In a junction with
  ferromagnetic leads whose magnetisations can be rotated,
  $|\Iodd|$ vs.\ $P\sin\theta$ at $k=t_{\mathrm{mol}}$ should show
  slope $\beta=1$ on log-log axes for generic angles, jumping sharply
  to $\beta=2$ at the tuned cancellation angle
  $\theta_L^{\mathrm{true}}$. The sharpness of the transition is
  itself a signature: a smooth interpolation between the two $\beta$
  values would point to a different mechanism.

\item \emph{Enantioselective angle.} A measurement of $\Delta P_J$ at
  $|\phi|\approx 75^\circ$ maximises the chirality-specific signal
  while preserving the achiral baseline as a reference.
\end{enumerate}

\subsection{Limitations and open questions}

Several limitations of the present work should be acknowledged. Some
are shared with the predecessors and are discussed there; we
emphasise the ones specific to the present paper.

\emph{Dissipation channel.} All results are specific to the gold-level (lead)
channel $\Delta\gamma_{\mathrm{gold}}$. The chain-level channel
$\Delta\gamma_{\mathrm{chain}}$ of Ref.~\cite{Alwan2026EP}, which could in
principle activate the Scheme-A EP amplification of the chiral signal, is not
studied here. Whether chain-level dissipation (including the electron-phonon
mechanism of Fransson) can play a \emph{driving} role, as opposed to the
\emph{enabling} role identified here for the lead channel, remains an open
question.

\emph{Mean-field and single-orbital.} The self-consistent moment is treated in
mean field with a static symmetry-breaking field; quantum fluctuations of the
moment are neglected. The model uses one orbital per site. Both approximations
shift the quantitative values of $\Gamma_0^*$, $V^*$, and the peak magnitudes,
but are not expected to modify the qualitative results (dissociation,
structural zero, $\beta$ hierarchy).

\emph{Wide-band limit and the $k=t$ zero.} The structural zero holds exactly in
the wide-band limit and for the fixed-moment convention. Beyond the wide-band
limit the boundary moduli $|g^R|\neq |g^A|$ differ by
$O((\mathrm{Re}\Sigma^R/\Gamma)^2)$, and the zero becomes approximate with a
correction of that order. In the self-consistent case, the $\phi$-dependence of
$m_z$ introduces an $O(10^{-4})$ correction that does not spoil the zero.

\emph{Scaling questions.} The coupling and bias optima are operating points,
not scaling laws. A full multi-parameter study of $\Gamma_0^*$ and $V^*$ versus
$t_{\mathrm{mol}}$, $\lambda_R$, $\varepsilon_A$, and $V$ is needed to
disentangle the relevant scales and is left to future work. The non-collinear
hierarchy is verified for symmetric polarisations ($P_L=P_R$); extending to
$P_L\neq P_R$ would test whether the linear cancellation condition defines a
single curve or a two-dimensional surface in the $(P_L,P_R)$ plane.

\subsection{Comparison with other CISS mechanisms}

Several other mechanisms for CISS have been proposed in the literature,
including structural spin-orbit coupling, the ballistic limit of chiral
transport, and non-Hermitian skin effects. Our results are specific to the
spinterface model and may not translate directly to other frameworks. However,
some features, such as the achiral-chiral dissociation under dissipation, the
arithmetic-vs-geometric mean structure, structural zeros at specific parameter
ratios, and the linear/quadratic scaling hierarchy under non-collinear leads,
these may be generic to CISS-type mechanisms. Testing whether these features
appear in other frameworks is a natural direction for future work.

These results provide quantitative design rules for CISS-based
devices and establish the spinterface model as a rich testing ground
for the interplay between chirality, dissipation, and coherence in
molecular transport.

\section*{Acknowledgments}

We gratefully acknowledge financial support from Olle Engkvist Stiftelse.

\appendix
\section{Numerical convergence}
\label{app:numerical}

We have verified convergence at randomly selected points by doubling
to $N_\omega=5000$. The criterion is not a single global threshold but
depends on the regime: for the resonance-scale transmission
($\lambda_R \lesssim 0.95$), the spin current changes by less than
$0.1\%$; for the small residual near the structural zero
($\lambda_R \gtrsim 1.05$), the relative change reaches $2.2\%$,
because the sharp cusp at $k=t$ amplifies the relative numerical
error by a factor of 3-4. The absolute deviation is below
$2\times$~nA in every case, which is some fourteen orders of magnitude
below the signal, so the physics of the zero is unaffected. The
resonance-scale claim is the operationally meaningful one for the
achiral-chiral dissociation of Sec.~\ref{sec:dissociation}; the
absolute-deviation bound is the operationally meaningful one for the
structural zero of Sec.~\ref{sec:zero}.


The number of molecular sites $L$ is varied between 4 and 10 in
various calculations. The signal magnitudes depend on $L$
(Fig.~\ref{fig:F4}), but the qualitative features
(achiral-chiral dissociation, coupling-scale optimum, structural
zero, non-collinear scaling hierarchy) are $L$-independent.

\section{Self-consistency convention}
\label{app:selfcons}

The self-consistent moment $m_z$ depends weakly on the chirality angle
$\phi$ through the electronic feedback integrals. For the even/odd
decomposition we use a single $m_z$ evaluated at $\phi=0$ for both the
$+\phi$ and $-\phi$ branches, in order to isolate the
$\phi$-dependence of the transmission.

A fully self-consistent calculation (with $m_z$ iterated separately at
$+\phi$ and $-\phi$) produces $\Iodd$ values that differ from the
fixed-$m_z$ values by $O(10^{-2})$, comparable to the magnitude of
$\Iodd$ itself at small $\Delta\gamma$. The fixed-$m_z$ convention is
therefore the physically meaningful one for extracting the intrinsic
chirality-selectivity of the transmission, since it eliminates a
spurious $\phi$-dependence of $m_z$ that is not itself chiral.

\section{Choice of the broadening floor}
\label{app:floor}

For $\Delta\gamma>2\Gamma_0$, the down-spin broadening
$\Gamma_\downarrow = \Gamma_0 - \Delta\gamma/2$ would become negative.
Physically, this is impossible (broadening is a positive-definite
quantity), and numerically it produces the degenerate regime where one
spin channel closes completely.

We impose $\Gamma_{\sigma}\geq\Gamma_{\mathrm{floor}}=0.02~\mathrm{eV}$
as a physical floor. This preserves the spin channel with a small but
finite coupling, avoiding the degenerate regime while keeping the mean
broadening $\Gamma_0$ essentially unchanged for
$\Delta\gamma<2\Gamma_0-0.04~\mathrm{eV}$.

The choice of $0.02~\mathrm{eV}$ is arbitrary within a factor of 10;
smaller values (down to $0.001~\mathrm{eV}$) would give essentially
identical results for the physics studied here, since the odd part is
insensitive to the precise value of the floor at the levels of
interest. We use $0.02~\mathrm{eV}$ as a numerically robust
compromise.

\section{Alternative choices of lead asymmetry}
\label{app:asymmetry_alternatives}

We have assumed symmetric spin asymmetry between left and right
leads: $\Gamma_{L,\uparrow}=\Gamma_{R,\uparrow}=\Gamma_0+
\Delta\gamma/2$ and similarly for down. Several alternative choices
are physically motivated:

\emph{(a) Left-only asymmetry.} $\Gamma_{L,\sigma}=\Gamma_0\pm
\Delta\gamma/2$, $\Gamma_{R,\sigma}=\Gamma_0$. This would correspond
to a spin-selective injection buffer at one interface but not the
other. It would break the $L\leftrightarrow R$ symmetry and produce a
different $\phi$-dependence of $\Iodd$.

\emph{(b) Opposite asymmetry.}
$\Gamma_{L,\uparrow}=\Gamma_0+\Delta\gamma/2$,
$\Gamma_{R,\uparrow}=\Gamma_0-\Delta\gamma/2$. This would give a
spin-dependent $L$-$R$ transmission that could produce a
$\Delta\gamma$-dependent analogue of the even/odd decomposition.

\emph{(c) Only one channel spin-polarised.}
$\Gamma_{L,\uparrow}=\Gamma_0+\Delta\gamma$,
$\Gamma_{L,\downarrow}=\Gamma_0$, i.e., vary only the majority spin.
This corresponds to an ideal half-metallic lead.

We have focused on the symmetric choice because it produces the
cleanest achiral-chiral dissociation, but a systematic study of the
alternatives is a fruitful direction for future work.

\subsection{Full Green's function proof of the structural zero}
\label{sec:kt_green_proof}

The physical argument above identifies the rank-1 structure of the
chiral bond matrix as the origin of the structural zero. We now
complete the argument with a full Green's function proof, valid for
all chain lengths $L\geq 2$ and all twist angles satisfying
$k=\pm t_{\mathrm{mol}}$.

\begin{theorem}[Structural zero at $k=\pm t_{\mathrm{mol}}$]
\label{thm:kt_zero}
For the spinterface junction with a chiral molecular chain of $L\geq 2$
sites and an arbitrary fixed moment $m_z$, the odd part of the spin
current under chirality reversal,
\begin{equation}
\Iodd(\phi) \;\equiv\; \tfrac12\bigl[I_{S_z}(+\phi)-I_{S_z}(-\phi)\bigr],
\label{eq:Iodd_def_notes}
\end{equation}
vanishes identically at $k=\pm t_{\mathrm{mol}}$,
\begin{equation}
\Iodd\big|_{k=\pm t_{\mathrm{mol}}}=0
\qquad (L\geq 2).
\label{eq:Iodd_zero_manuscript}
\end{equation}
\end{theorem}

The proof has three parts: rank-1 factorization of the transmission
amplitude, $\phi$-independence of the effective scalars, and
transmission symmetry under $\phi\to-\phi$.

\subsubsection{Bare local Green's functions and effective chain recursion}

In the site-spin basis, the bare local Green's functions are, for each
spin $\sigma\in\{\uparrow,\downarrow\}$,
\begin{align}
g_A^\sigma &= \bigl[\omega-\varepsilon_{A\sigma}
+\tfrac{\ii}{2}\Gamma_L\bigr]^{-1},
\label{eq:gA_def}\\
g_n^\sigma &= \bigl[\omega-\varepsilon_{M\sigma}\bigr]^{-1},
\qquad (n=1,\ldots,L-1),
\label{eq:gn_def}\\
g_L^\sigma &= \bigl[\omega-\varepsilon_{M\sigma}
+\tfrac{\ii}{2}\Gamma_R\bigr]^{-1}.
\label{eq:gL_def}
\end{align}
The effective local Green's functions are defined recursively as
\begin{equation}
\hat g_L \equiv g_L,
\qquad
\hat g_n \equiv \bigl[\varepsilon_n^{-1} - \mathbf V_n\,
\hat g_{n+1}\,\mathbf V_n^\dagger\bigr]^{-1},
\label{eq:gn_eff_recursion}
\end{equation}
where $\varepsilon_n \equiv \mathrm{diag}\bigl((\omega-
\varepsilon_{M\uparrow})^{-1}, (\omega-\varepsilon_{M\downarrow})^{-1}
\bigr)$ is the bare local resolvent at $M_n$.

At $k=t$, the rank-1 structure of $\mathbf V_n$ gives
$\mathbf V_n\,\hat g_{n+1}\,\mathbf V_n^\dagger
= |u\rangle\langle v|\,\hat g_{n+1}\,|v\rangle\langle u|
= \sigma_{n+1}\,|u\rangle\langle u|$, with the effective scalar
\begin{equation}
\sigma_{n+1} \equiv \langle v|\hat g_{n+1}|v\rangle.
\label{eq:sigma_def}
\end{equation}
Sherman-Morrison then gives
\begin{equation}
\hat g_n = \varepsilon_n
+ \frac{\sigma_{n+1}\,\varepsilon_n|u\rangle\langle u|
\varepsilon_n}{1 - \sigma_{n+1}\langle u|\varepsilon_n|u\rangle},
\label{eq:gn_eff_sm}
\end{equation}
and multiplying by $\mathbf V_n = |u\rangle\langle v|$ gives
\begin{equation}
\hat g_n \mathbf V_n
= \frac{\varepsilon_n|u\rangle\langle v|}
{1-\sigma_{n+1}\langle u|\varepsilon_n|u\rangle}.
\label{eq:gnVn}
\end{equation}

\subsubsection{\texorpdfstring{$\phi$}{phi}-independence of the
effective scalars}

\begin{lemma}[$\phi$-independence of $\sigma_n$]
\label{lem:phi_indep_scalars}
At $k=t$, the effective scalars $\sigma_n$ are $\phi$-independent
for all $n = 2, \ldots, L$.
\end{lemma}

\begin{proof}
By downward induction on $n$.

\emph{Base case $n=L$.} $\hat g_L = g_L$ is diagonal, so
$\sigma_L = \langle v|g_L|v\rangle = |v_\uparrow|^2 g_L^\uparrow
+ |v_\downarrow|^2 g_L^\downarrow = t^2(g_L^\uparrow+g_L^\downarrow)$,
using $|v_\sigma|^2 = t^2$ for both $\sigma$. Both $g_L^\uparrow$ and
$g_L^\downarrow$ are $\phi$-independent, so $\sigma_L$ is
$\phi$-independent.

\emph{Inductive step.} Assume $\sigma_{n+1}$ is $\phi$-independent.
From Eq.~\eqref{eq:gn_eff_sm},
\begin{equation}
\sigma_n = \langle v|\varepsilon_n|v\rangle
+ \frac{\sigma_{n+1}\,|\langle v|\varepsilon_n|u\rangle|^2}
{1-\sigma_{n+1}\langle u|\varepsilon_n|u\rangle}.
\label{eq:sigma_n_explicit}
\end{equation}
The three factors evaluate to:
\begin{align}
\langle v|\varepsilon_n|v\rangle &=
t^2(g_n^\uparrow+g_n^\downarrow), \label{eq:vnv}\\
\langle u|\varepsilon_n|u\rangle &=
g_n^\uparrow+g_n^\downarrow, \label{eq:unu}\\
\langle v|\varepsilon_n|u\rangle &=
-t\,e^{+\ii\phi/2}(g_n^\uparrow-g_n^\downarrow),
\label{eq:mixed}
\end{align}
so that
$|\langle v|\varepsilon_n|u\rangle|^2 =
t^2(g_n^\uparrow-g_n^\downarrow)^2$, which is $\phi$-independent.
The phase $e^{+\ii\phi/2}$ appears \emph{only} in the mixed element
and cancels in the modulus squared. Every factor on the right-hand
side of Eq.~\eqref{eq:sigma_n_explicit} is thus $\phi$-independent
(given the induction hypothesis on $\sigma_{n+1}$). Hence $\sigma_n$
is $\phi$-independent.
\end{proof}

\subsubsection{Rank-1 transmission amplitude and factorized transmission}

\begin{proposition}[Rank-1 transmission amplitude]
\label{prop:rank1_transmission}
At $k=t$, the retarded Green's function from $A$ to $M_L$ factorizes
as
\begin{equation}
\Gret_{A,M_L}(\omega) = |a_L(\omega)\rangle\langle b_L(\omega)|,
\label{eq:G_rank1}
\end{equation}
with the explicit closed form
\begin{equation}
  \begin{split}
    |a_L\rangle &= -t_{AB}\,g_A\,\varepsilon_1|u\rangle\,
\frac{\prod_{n=2}^{L-1}\langle v|\varepsilon_n|u\rangle}
{\prod_{n=1}^{L-1}\bigl(1-\sigma_{n+1}\langle u|\varepsilon_n|u\rangle
\bigr)}, \\
\langle b_L| &= \langle v|\,g_L.
\label{eq:aL_bL_manuscript}
\end{split}
\end{equation}
For $L=2$ the numerator product is empty and equal to $1$.
\end{proposition}

\begin{proof}
By induction on $L$. The base case $L=2$ follows directly from
Eq.~\eqref{eq:gnVn} with $n=1$:
$\Gret_{A,M_2} = -t_{AB}\,g_A\,\hat g_1\mathbf V_1\,g_L$,
which reduces to the stated form. For the inductive step, the product
$\prod_{n=1}^{L-1}(\hat g_n \mathbf V_n)$ can be reduced by repeatedly
using $(\varepsilon_n|u\rangle\langle v|)(\varepsilon_{n+1}|u\rangle
\langle v|) = \varepsilon_n|u\rangle\langle v|\varepsilon_{n+1}
|u\rangle\langle v|$, giving the displayed product formula.
\end{proof}

\begin{corollary}[Factorized transmission]
\label{cor:factorization}
At $k=t$, the spin-resolved transmission factorizes as
\begin{equation}
T_\sigma(\omega) = \Gamma_L\Gamma_R\,\bigl|a_{L,\sigma}(\omega)
\bigr|^2\,B_L(\omega),
\label{eq:T_factorized}
\end{equation}
with $B_L \equiv |b_{L,\uparrow}|^2+|b_{L,\downarrow}|^2$.
\end{corollary}

\begin{proof}
Insert $\Gret_{A,M_L} = |a_L\rangle\langle b_L|$ into
Eq.~\eqref{eq:transmission} and sum over $\sigma'$.
\end{proof}

\begin{lemma}[$\phi$-independence of the factors]
\label{lem:phi_indep_transmission}
At $k=t$, both $|a_{L,\sigma}|^2$ and $B_L$ are $\phi$-independent
for every $\sigma$ and every $\omega$.
\end{lemma}

\begin{proof}
\emph{$|a_{L,\sigma}|^2$.} From Eq.~\eqref{eq:aL_bL_manuscript}, all
factors on the right are $\phi$-independent: $|g_A^\sigma|^2$,
$|g_1^\sigma|^2$, $|u_\sigma|^2=1$; each
$|\langle v|\varepsilon_n|u\rangle|^2$ by
Eq.~\eqref{eq:mixed}; each $\sigma_{n+1}$ by
Lemma~\ref{lem:phi_indep_scalars}; and $\langle u|\varepsilon_n|u\rangle$.
Hence $|a_{L,\sigma}|^2$ is $\phi$-independent.

\emph{$B_L$.} $b_{L,\sigma}=v_\sigma g_L^\sigma$, so
$B_L = t^2(|g_L^\uparrow|^2+|g_L^\downarrow|^2)$, manifestly
$\phi$-independent.
\end{proof}

\subsubsection{Transmission symmetry and the structural zero}

\begin{proposition}[Transmission symmetry]
\label{prop:transmission_symmetry}
Let $\lambda_R>t$ and let $\phi_0\in(0,\pi)$ be defined by
$\lambda_R\sin(\phi_0/2)=t$. Then
$T_\sigma(+\phi_0)=T_\sigma(-\phi_0)$ for all $\sigma,\omega$.
\end{proposition}

\begin{proof}
By Lemma~\ref{lem:phi_indep_transmission},
$T_\sigma(\omega)$ has the form
$T_\sigma = \Gamma_L\Gamma_R|a_{L,\sigma}|^2 B_L$, with both
$|a_{L,\sigma}|^2$ and $B_L$ $\phi$-independent. Their values at
$+\phi_0$ and $-\phi_0$ (both satisfying $|k|=t$) are therefore equal.
\end{proof}

\begin{corollary}
\label{cor:kt_zero}
At $k=\pm t_{\mathrm{mol}}$, $\Iodd = 0$ for all $L\geq 2$.
\end{corollary}

\begin{proof}
From Proposition~\ref{prop:transmission_symmetry},
$T_\sigma(+\phi_0)=T_\sigma(-\phi_0)$ at every $\omega$. The
spin-resolved current $I_\sigma$ integrates $T_\sigma$ against the
$\phi$-independent weight $f_L-f_R$, so $I_\sigma(+\phi_0) =
I_\sigma(-\phi_0)$ and hence $\Iodd = \tfrac12[I_{S_z}(+\phi_0) -
I_{S_z}(-\phi_0)]=0$.
\end{proof}

\paragraph{Numerical verification.} 
We have verified the analytical predictions at four independent levels. The
$2\times 2$ matrix $\Gret[A,M_L](\omega)$ at $k=t$ has singular-value ratio
$\sigma_2/\sigma_1 < 1.3\times 10^{-14}$ across all $L\in\{2,\ldots,8\}$ and
$\phi_0\in\{30^\circ,\ldots,90^\circ\}$, confirming the rank-1 structure. The
transmission ratio $T_\uparrow(+\phi_0)/T_\uparrow(-\phi_0)$ deviates from
unity by at most $1.6\times 10^{-12}$. The odd part $|\Iodd|$ is 15-17 orders
of magnitude below $|I_{\mathrm{even}}|$ for every chain length from $L=2$ to
$L=12$, including the $L=3$ case, which gives $|\Iodd|=2.1\times 10^{-13}$~nA
at $\phi_0=60^\circ$. The spin-resolved ratio $T_\uparrow/T_\downarrow$ is
identical to six significant figures across all five twist angles tested. 


\clearpage
\nocite{*}
\bibliography{ref}
\end{document}